\documentclass[structabstract]{aa}  
\usepackage{graphicx}
\usepackage{txfonts}
\usepackage{natbib} 
\usepackage{url} 
\usepackage{subfig} 
\usepackage{multirow} 
\newcommand{\Msun}{$M_\odot$}
\newcommand{\Rsun}{$R_\odot$}

\begin{document}
\title{Observational properties of low redshift pair instability supernovae}

\author{A.~Kozyreva\inst{1}
\and
S.~Blinnikov\inst{2,3}
\and
N.~Langer\inst{1}
\and
S.-C.~Yoon\inst{4}
}
\institute{
Argelander-Institut f\"ur Astronomie, Universit\"at Bonn, Auf dem H\"ugel 71, 53121 Bonn, Germany \\
\email{kozyreva@astro.uni-bonn.de}
\and
Institute for Theoretical and Experimental Physics, Bolshaya Cheremushkinskaya Str. 25, 117218 Moscow, Russia 
\and
Kavli Institute for the Physics and Mathematics of the Universe (WPI), The University of Tokyo, 5-1-5 Kashiwanoha, Kashiwa, Chiba
277-8583, Japan
\and
Department of Physics \& Astronomy, Seoul National University, Gwanak-ro 1, Gwanak-gu, 151-742, Seoul, South Korea
}


 
\abstract
{  
So called superluminous supernovae have been recently discovered in the local Universe.  
It appears possible that some of them originate from
stellar explosions induced by the pair instability mechanism.  
Recent stellar evolution models also predict pair instability supernovae
from very massive stars at fairly high metallicities (i.e., $Z \sim 0.004$). 
}
{
We provide supernova models and synthetic light curves for two progenitor models,
a 150~\Msun{} red-supergiant and a 250~\Msun{} yellow-supergiant at a 
metallicity of $Z = 0.001$, for which the
evolution from the main sequence to collapse, and the initiation of the 
pair instability supernova (PISN) itself, has been previously computed 
in a realistic and self-consistent way.
}
{
We are using the radiation hydrodynamics code STELLA to describe the supernova evolution
of both models over a time frame of about 500 days.
}
{
We describe the shock-breakout phases of both supernovae 
which are characterized by a higher luminosity, a longer duration and a lower effective temperature
than those of ordinary Type~IIP supernovae.
We derive the bolometric as well as the \emph{U}, \emph{B}, \emph{V}, \emph{R} and 
\emph{I} light curves of our pair instability supernova models,
which show a long-lasting plateau phase with maxima at $M_{\rm bol}\simeq -19.3$~mag and $-21.3$~mag for our
lower and higher mass model, respectively. 
While we do not produce synthetic spectra,
we also describe the photospheric composition and velocity as function of time.
}
{
We conclude that the light curve of the explosion of our initially 150~\Msun{} star
resembles those of relatively bright type~IIP supernovae, whereas its photospheric velocity at early times
is somewhat smaller. Its $^{56}$Ni mass of 0.04~\Msun{} also falls well into the range found in ordinary core collapse
supernovae. The light curve and photospheric velocity of our 250~\Msun{} models has a striking resemblance
with that of the superluminous SN~2007bi, strengthening its interpretation as pair instability supernova.   
We conclude that pair instability supernovae may occur more frequently in the local universe than previously assumed.
}


\maketitle

\section[Introduction]{Introduction}
\label{sect:intro2}

The final fate of very massive stars with initial masses between approximately
140~\Msun{} and 260~\Msun{} has been studied in many papers
\citep{1964ApJS....9..201F,1967SvA....10..604B,1967ApJ...148..803R,1967PhRvL..18..379B,1968Ap&SS...2...96F,1986A&A...167..274E,2003ApJ...591..288H}.  
Such massive stars undergo the dynamical instability caused by the creation of
electron-positron pairs in oxygen cores, if they can retain their oxygen core masses
high enough ($\gtrsim 60$~\Msun{}) until carbon exhaustion at the center \citep{2003ApJ...591..288H}.  
This leads to explosive oxygen burning that eventually causes complete
disruption of the star without leaving a compact remnant behind.  

Theoretical models predict that these pair instability supernovae (PISNe) can be
much more energetic and luminous than ordinary SNe.  Explosion energies 
of up to $10^{\,53}~\mathrm{erg}$ and masses of radioactive nickel up to
40~\Msun{}~\citep[e.g.,][]{2002ApJ...567..532H} are found to be achieved depending on the progenitor
mass.  The corresponding light curves are characterized by a long-duration of
several hundreds of days and luminosities of up to $10^{\,43} -
10^{\,44}~\mathrm{erg~s^{-1}}$
\citep{2005ApJ...633.1031S,2011ApJ...734..102K,2013MNRAS.428.3227D,2013ApJ...777..110W,2013ApJ...762L...6W}.

It is believed that PISNe are particularly relevant to the first generations of
stars in the early Universe.  Theoretical studies indicate that a significant
fraction of the first stars would be massive enough to be potential progenitors
of PISNe, mainly because of the lack of efficient coolants in the star-forming
regions in the early Universe
\citep[e.g.,][]{1999ApJ...527L...5B,2001ApJ...548...19N,2002Sci...295...93A,2003ApJ...589..677O,
2006ApJ...648...31O, 2009ApJ...706.1184O}.  Such very massive stars in the
early Universe  would not lose much mass during the pre-supernova evolutionary
stages in favor of PISN production, because metal-free massive stars are not
supposed to have strong line-driven winds \citep{2006A&A...446.1039K} and
because they are expected to be stable against pulsations
\citep{2001ApJ...550..890B}.  Several numerical studies have been therefore
presented to discuss the nature and detectability of PISNe in the early
Universe \citep{2005ApJ...633.1031S,
2011ApJ...734..102K,2013MNRAS.428.3227D,2013ApJ...777..110W}.  These studies
considered a variety of PISN progenitors including red supergiants (RSG),
blue supergiants (BSG) and pure helium stars, but the considered metallicities
of these progenitor models were limited to zero or very small values ($Z \le
10^{\,-4}$).

\begin{table*}
\caption[Characteristics of the PISN models.]
{Characteristics of the PISN models: name of the model,
initial metallicity $Z$, 
initial/final mass ($M_{\mathrm{ini}}/M_{\mathrm{f}}$) in solar masses, 
radius of the star at the onset of the radiative
calculation $R$ in solar radii, explosion 
($E_{\mathrm{expl}}$) and kinetic ($E_{\mathrm{kin}}$) energy in foe ($=10^{\,51}$~erg), 
specific energy ($E/M$) in units [$10^{\,50}$~erg/$M_\odot${}],
the velocity at the outer edge of the $^{56}$Ni-rich layer in km~s$^{-1}$ ($v_\mathrm{ni,max}$), 
bulk yields of the isotopes in the ejecta in solar masses
(hydrogen $^1$H, helium $^4$He, carbon $^{12}$C, oxygen $^{16}$O, silicon $^{28}$Si, nickel $^{56}$Ni).
The names of the models starting with 'R' indicate red supergiants. 
The names starting with `B' are blue supergiants.
Models labeled `.D'
\citep{2013MNRAS.428.3227D} and labeled `.K'
\citep{2011ApJ...734..102K} are given for comparison.}
\label{table:moddata}
\begin{center}
\begin{tabular}{rccccccccccccc}
\hline
Name & $Z$ & $M_{\mathrm{in}}/M_{\mathrm{f}}$ & $R$ & $E_{\mathrm{expl}}$ & $E_{\mathrm{kin}}$
& $E/M$ & $v_\mathrm{ni,max}$ & $^1$H & $^4$He & $^{12}$C & $^{16}$O & $^{28}$Si & $^{56}$Ni \\
\, &  & ($M_\odot$) & $(R_\odot)$ & (foe) & (foe) & & (km/s) & ($M_\odot$) & ($M_\odot$) & ($M_\odot$) & ($M_\odot$) & ($M_\odot$)
& ($M_\odot$) \\
\hline
\textbf{150M} & $10^{\,-3}$        & 150/94   & 3394  & 12       & 8     & 1.3 &  500    &  5   & 24  & 2  & 47 & 6      & 0.04 \\
R150.K        & $2\times10^{\,-6}$ & 150/143  & 2314  & 9        &       & 0.6 & no data &  50  & 21  &    &    &        & 0.07 \\
R190.D        & $2\times10^{\,-6}$ & 190/164  & 4044  & 44       & 33    & 2.7 & 1800    &  24  & 46  & 5  & 78 & 0.05   & 2.63 \\
R250.K        & $2\times10^{\,-6}$ & 250/236  & 3214  & 69       &       & 2.9 & 5000    &  73  & 39  &    &    &        & 37.9 \\
\hline
B190.D        & $2\times10^{\,-6}$ & 190/134  &  186  & $\sim$44 & 34    & 3.3 & 1200 &   6   & 34  & 5 & 78 & 0.05   & 2.99 \\
B210.D        & $2\times10^{\,-6}$ & 210/147  &  146  & 75       & 66    & 5.1 & 4400 &  4    & 31  & 6 & 93 & 0.06   & 21.3 \\
\textbf{250M} & $10^{\,-3}$        & 250/169  &  745  & 70       & 44    & 4.1 & 5000 &  10   & 48  & 1 & 42 & 23     & 19.3 \\
B250.K        & $0$                & 250/250  &  187  & 63       &       & 2.5 & 3200 &  86   & 40  &   &    &        & 23.1 \\
\hline
\end{tabular}
\end{center} 
\end{table*}

However, several recent studies indicate that PISNe are likely to occur not
only in the early Universe, but also in the local Universe where metallicity is
systematically higher than in the environments at high redshift.  
\citet{2010MNRAS.408..731C} found several very massive stars (VMS) with initial
masses of $150~M_\odot \lesssim M \lesssim 320~M_\odot$ in
the Large Magellanic Clouds (LMC), which are potential PISN progenitors. The
final fate of such VMS stars is critically determined by mass loss
\citep[e.g.,][]{2011A&A...531A.132V}.  Given the strong metallicity dependence
of the stellar winds mass loss rate \citep[e.g.,][]{2007A&A...473..603M}, it is
generally believed that VMSs can not retain enough mass to produce PISNe for
high metallicity.  However, \citet{2007A&A...475L..19L} point out that the production
of PISNe does not necessarily require extremely metal poor environments,
although low metallicity is still preferred.  Using stellar evolution models,
they argue that the metallicity threshold below which PISNe may occur
($Z_\mathrm{PISN}$) can be as high as $Z_\odot/3${} \citep{2009Natur.462..579L}.  More
recently  \citet{2013MNRAS.433.1114Y} drew a similar conclusion.

The recent discovery of several superluminous SNe (SLSNe)  that cannot be
easily explained by usual core-collapse and/or interaction supernovae
\citep[see][for a recent review]{2012Sci...337..927G} also provides evidence for
PISN in the local Universe.  In particular, the observed properties of SN 2007bi and SN~2213-1745
seem to imply a large amount of radioactive nickel in these supernovae  (i.e, more than 3~\Msun{}
of $^{56}$Ni), for which PISN explosion gives
one of the best explanations 
\citep{2009Natur.462..624G,2010A&A...512A..70Y,2012Natur.491..228C}.  
Given that some other possibilities like very energetic core-collapse explosion
or magnetar-driven SN have also been suggested to explain SLSNe
\citep{2010ApJ...717L..83M, 2010ApJ...719L.204W, 2010ApJ...717..245K} and that future observational
SN surveys will discover more diverse SLSN events in nearby galaxies, detailed
studies on the observable properties of PISNe in the local Universe are still
needed to have a solid conclusion on the association of SLSNe and PISNe
\citep[e.g.,][]{2012MNRAS.426L..76D}.

\citet{2007A&A...475L..19L} presented PISN progenitor models with initial
masses of 150~\Msun{} and 250~\Msun{}, which roughly represent the low- and
high-mass ends of PISN progenitors, at a metallicity of {\emph Z}=0.001 to discuss the
possible event rate of PISNe in the local Universe.  \citet{2014paper1..K}
investigated the explosions and consequent nucleosynthesis of these models to
discuss the implications of PISNe for the chemical evolution.  In this paper,
we explore these models further to investigate their observable properties
using the radiation hydrodynamics code {\sc{
STELLA}}~\citep{2006A&A...453..229B}.  These models together with those from
\citet{1990A&A...233..462H} and \citet{2013arXiv1312.5360W} are among the highest metallicity
PISN light curve models available in the literature, and therefore useful
to identify PISN events at low redshift.

We explain the method of calculations in Section~\ref{sect:method2} and discuss the produced light curves in 
Section~\ref{sect:results2}.  In Section~\ref{sect:discussion2}, we compare our results with other
synthetic light curves for PISNe and with observational light curves of usual and unusual 
core collapse SNe (CCSNe).  We conclude our work in
Section~\ref{sect:conclusions2}.

\section[Evolutionary models and light curves modeling]{Evolutionary models and light curves modeling}
\label{sect:method2}

\subsection[Description of the evolutionary models]{Description of the evolutionary models}
\label{subsect:evol}

For simulating the light curves of PISNe we use the results of
evolutionary calculations produced with the {\sc Binary Evolution Code} ({\sc{BEC}},
\citealt{2007A&A...475L..19L,2014paper1..K}).  These are initially
150~$M_\odot${} and 250~$M_\odot${} models at a metallicity of $Z=10^{\,-3}$, with an
initial rotation velocity of $10~\mathrm{km~s^{-1}}$.  The evolution of the stars was
calculated all the way from the zero age main sequence to the
thermonuclear explosion caused by the pair instability.  
To follow the nucleosynthesis, \citet{2014paper1..K} used the Torch nuclear
network developed by \citet{Timmes..public..networks,1999ApJS..124..241T} using
200 isotopes.  Our progenitor models are red and yellow supergiants 
for the 150~\Msun{} and 250~\Msun{} models, respectively
\citep{2007A&A...475L..19L,2014paper1..K}.  The main characteristics of the
models are summarized in Table~\ref{table:moddata}.

\begin{figure}
\centering
\includegraphics[width=0.5\textwidth]{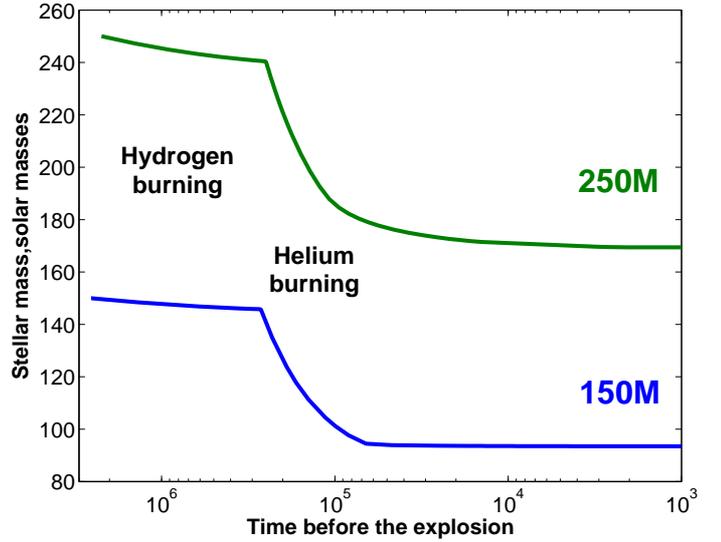}
\caption[The pre-supernova evolution of the stellar mass of our models due to stellar wind mass loss.]
{The pre-supernova evolution of the stellar mass of our models due to stellar wind mass loss.}
\label{figure:massloss}
\end{figure}

Our stellar models lose a large fraction of their initial mass during core hydrogen and core helium burning due to line-driven
winds (\citealt[see][for details]{2007A&A...475L..19L} and
\citealt[][]{1989A&A...219..205K,2001A&A...369..574V,2005A&A...442..587V,2010ApJ...725..940Y}).  
The average mass-loss rate is about $2\times 10^{\,-5}~M_\odot$~yr$^{\,-1}${}.  By the onset of the pair instability our 150~\Msun{} star
has lost 56~\Msun{} and our 250~\Msun{} star 81~\Msun{}.  
We show the pre-supernova evolution of stellar mass in Figure~\ref{figure:massloss}.

The stellar evolutionary models were calculated with the assumption of semi-convection using a large semi-convective mixing parameter
($\alpha_{\mathrm{SEM}}=1$), and without convective core overshooting.  The mixing length parameter
was chosen to be 1.5 \citep{2006A&A...460..199Y}.  Note that we neglected any convective mixing during the pair instability phase
\citep{2014paper1..K}.

The input models for the {\sc STELLA} calculations already contain the shock wave
generated at the boundary between the oxygen and helium shells emerging from
the interaction between the rapidly expanding inner core and the nearly static
outer helium-hydrogen envelope~\citep{2014paper1..K}.  The physics of the pair instability
explosion mechanism is well understood and smoothly reproduced by evolutionary
simulations without the need of invoking artificial assumptions.  This is
contrasted to other SN types that often need free parameters for
supernova modeling, such as the critical density for the transition from
deflagration to detonation for type~Ia~supernovae and the mass cut, degree of
chemical mixing,  and explosion energy for core-collapse supernovae. 

\begin{figure}
\centering
\includegraphics[width=0.5\textwidth]{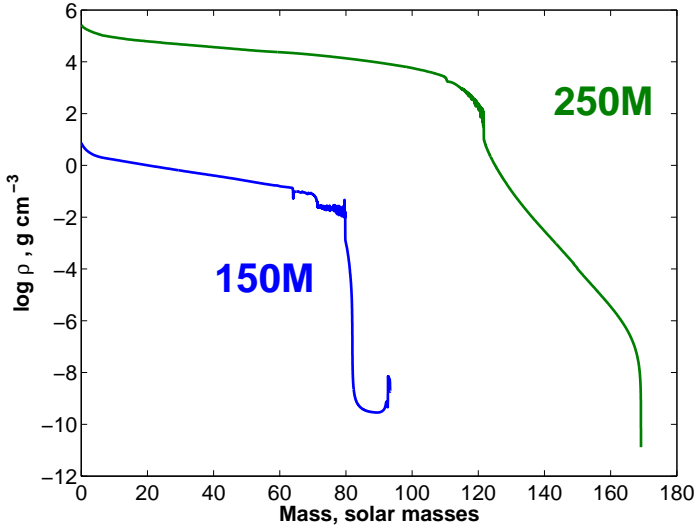}
\caption[The density structure of the 150~$M_\odot${} and 250~$M_\odot${} 
PISN progenitor models at the onset of the {\sc STELLA} calculations.]
{The density structure of the 150~$M_\odot${} and 250~$M_\odot${} PISN
progenitor models at the onset of the {\sc STELLA} calculations.}
\label{figure:rho}
\end{figure}

\begin{figure*}
\centering
{\includegraphics[width=0.5\textwidth]{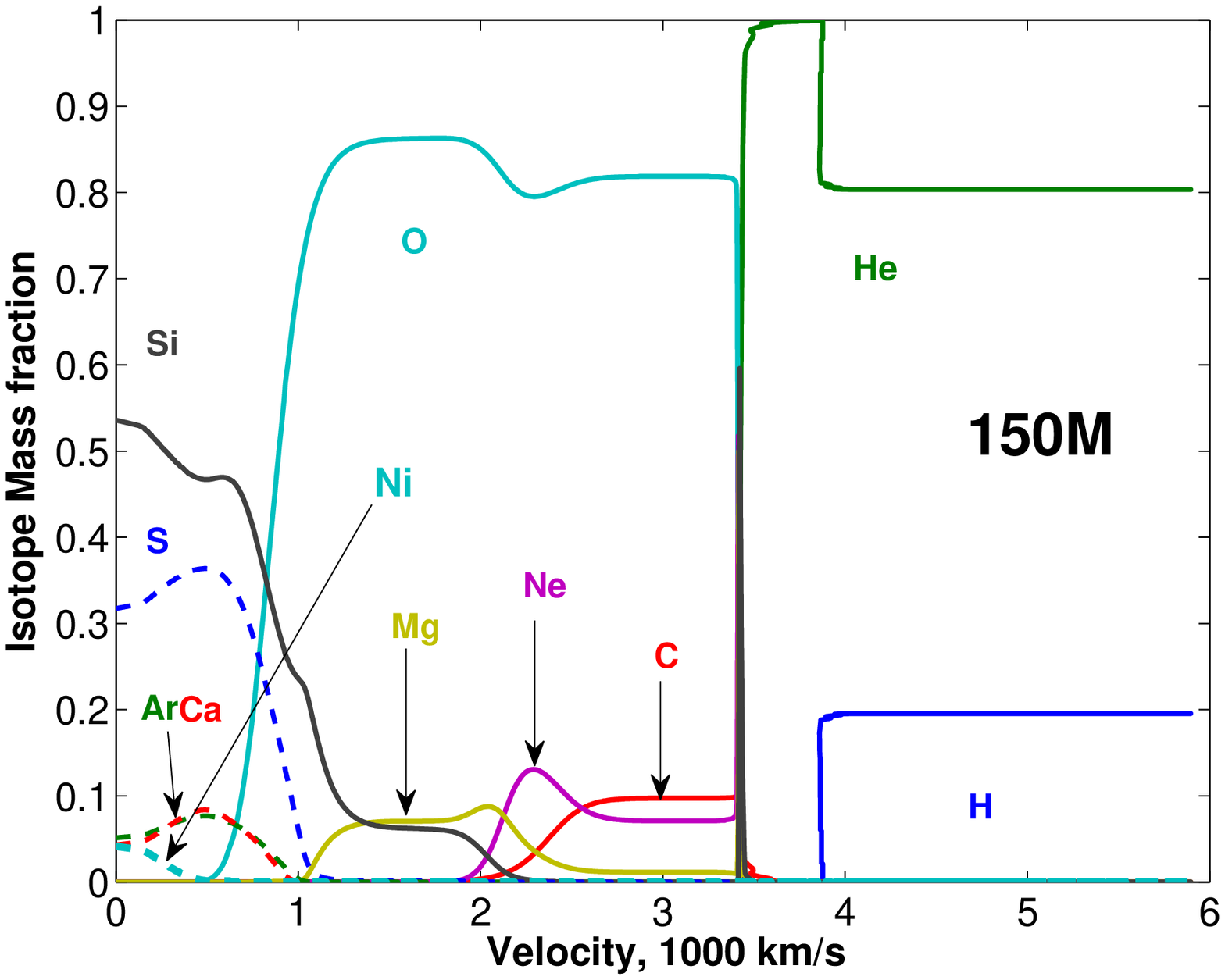}}~~
{\includegraphics[width=0.5\textwidth]{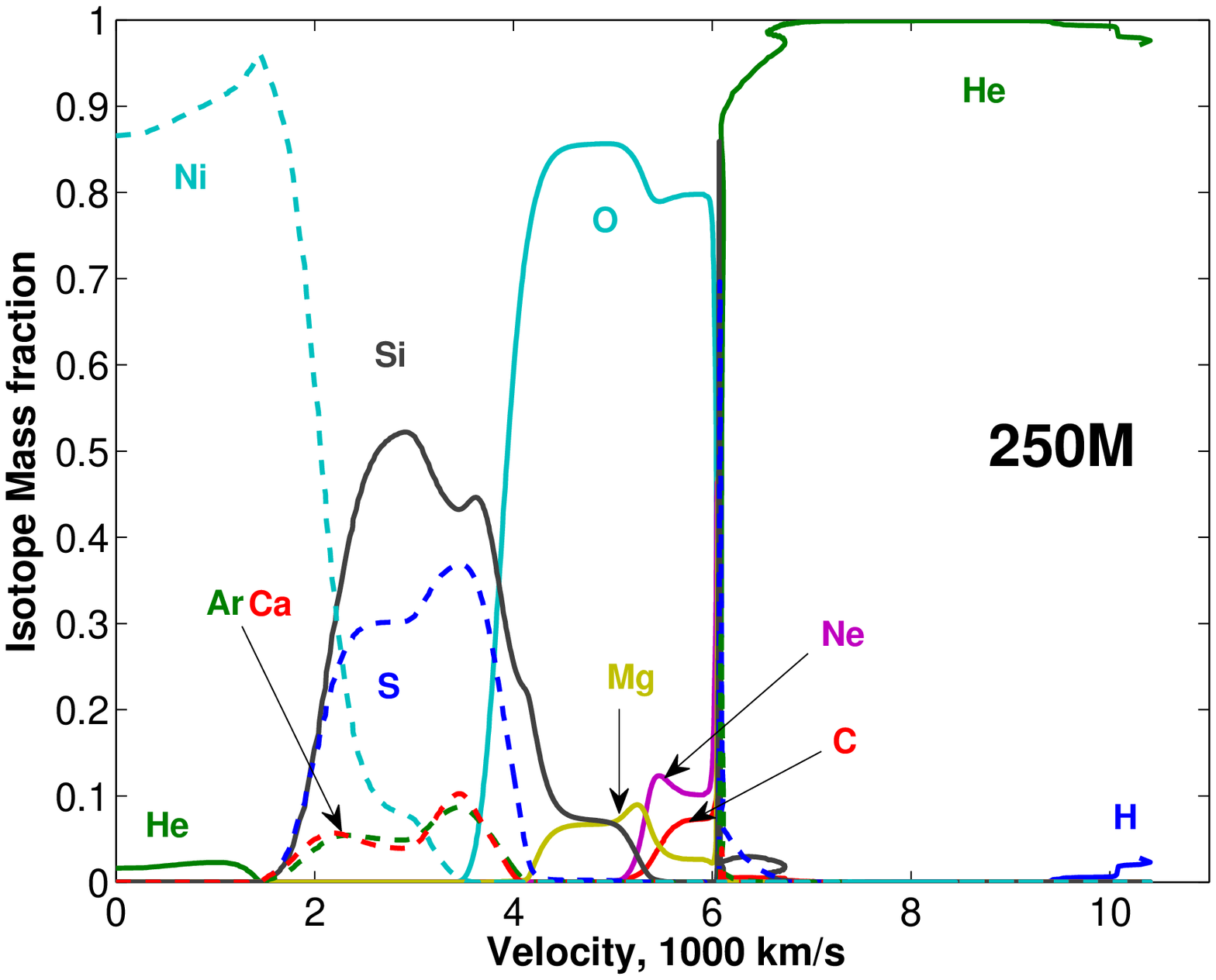}}
\caption[The chemical structure of the exploding 150~$M_\odot${} 
and 250~$M_\odot${} stars at metallicity $Z = 10^{\,-3}$ at the onset of the {\sc STELLA} calculations.]
{The chemical structure of the exploding 150~$M_\odot${} and 250~$M_\odot${}
stars at metallicity $Z = 10^{\,-3}$ at the onset of the {\sc STELLA} calculations.}
\label{figure:isostructure}
\end{figure*}

In Figures~\ref{figure:rho} and \ref{figure:isostructure}, we show the
density and the chemical structure of our progenitor models at the onset of the
{\sc STELLA} calculations.  At this moment the matter below the helium shell
expands nearly homologously ($v\propto r$).  In
Figure~\ref{figure:isostructure}, we truncated the outer hydrogen-helium
envelope which moves at very low velocity compared to the fast moving inner ejecta.  We also do not show in this figure the
bottom of the hydrogen-helium envelope where the shock is located and causes a strong discontinuities.  
The ejecta in the 150~\Msun{} model contain 0.04~\Msun{} of \,$^{56}$Ni in the
innermost region, moving at a velocity of less than 500~km~s$^{-1}$.  The 250~\Msun{} model
contains 19~\Msun{} of nickel over an extended region with $0~\mathrm{km~s^{-1}} \le v
\le 5000~\mathrm{km~s^{-1}}$, which corresponds to $0~M_\odot \le M_r
\le 50~M_\odot${}.  We do not apply any mixing in our models.  
Later we discuss the impact of this chemical
structure on light curves.  There is a small amount of helium in the innermost
region of the 250~\Msun{} star as seen in Figure~\ref{figure:isostructure}.  
This helium was produced by photo-disintegration of heavy elements during the
explosive burning.  We summarize the main properties of the PISNe in
Table~\ref{table:moddata} along with other PISN models that we use for
comparison.

\subsection[Simulation of light curves and SEDs]
{Simulation of theoretical light curves and SEDs\footnote{SED --- spectral energy distribution.}}
\label{subsect:LC}

For simulating the hydrodynamic evolution during the pair instability explosion we use 
the one-dimensional (1D) multigroup radiation Lagrangian implicit hydrodynamics code {\sc STELLA}
(\citealt{2006A&A...453..229B} and references therein).  
The code solves the hydrodynamic equations coupled with the radiative transfer equations without assuming radiative equilibrium.  
The {\sc STELLA} code uses one temperature for the matter and has no specific temperature for radiation.  
The non-steady radiative transfer is solved for each of the one hundred frequency groups in all radial zones.  
The energy groups are uniformly distributed in logarithmic scale (i.e. in geometric progression) 
between the maximum wavelength of 50,000~\AA{} and the minimum wavelength of 1~\AA{}.  
The light curves are computed by integration of fluxes calculated with the STELLA code with {\emph{UBVRI}} 
Bessel filter function using logarithmic interpolation.

\begin{figure*}
\centering
{\label{figure:bolbands150}\includegraphics[width=0.5\textwidth]{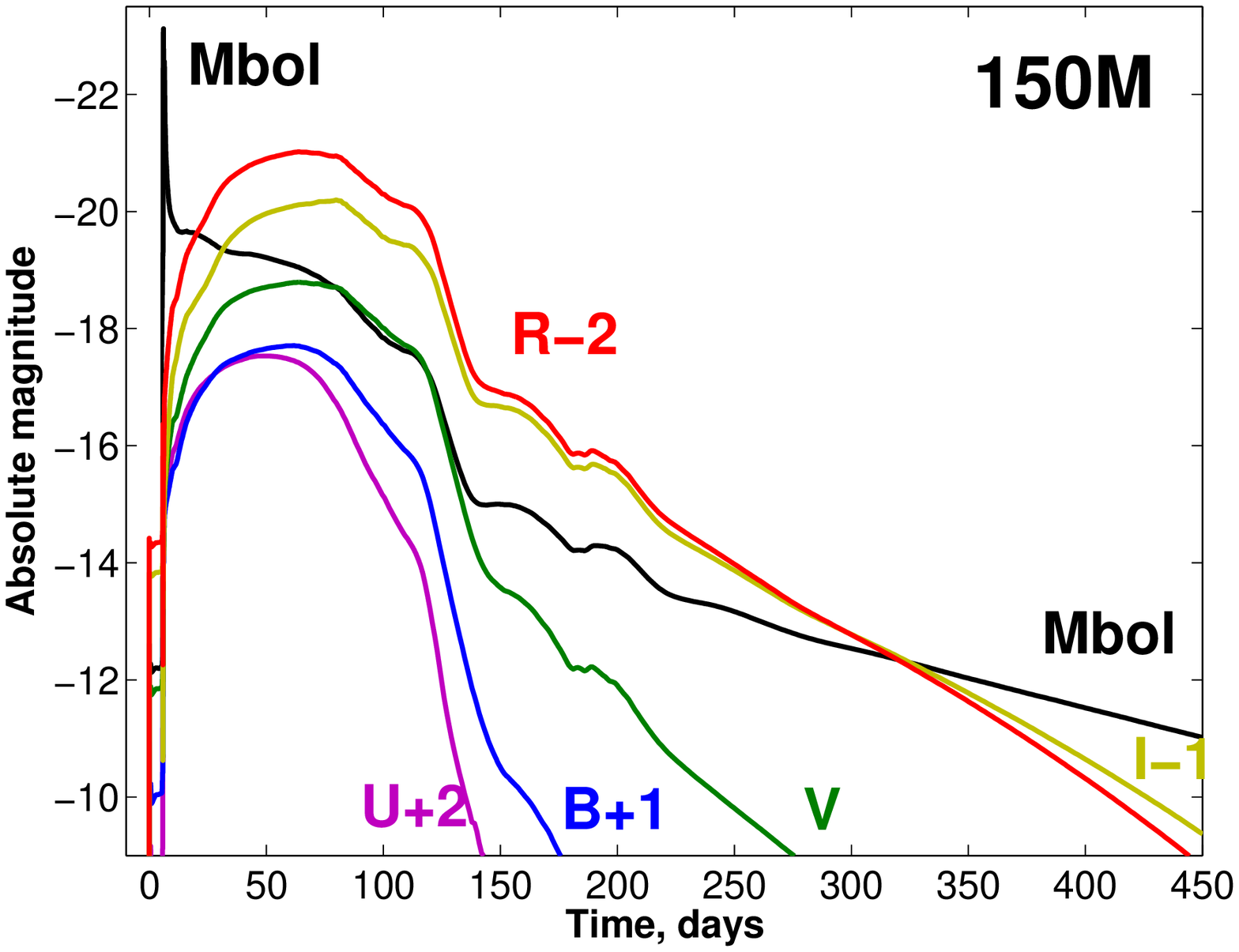}}~~
{\label{figure:bolbands250}\includegraphics[width=0.5\textwidth]{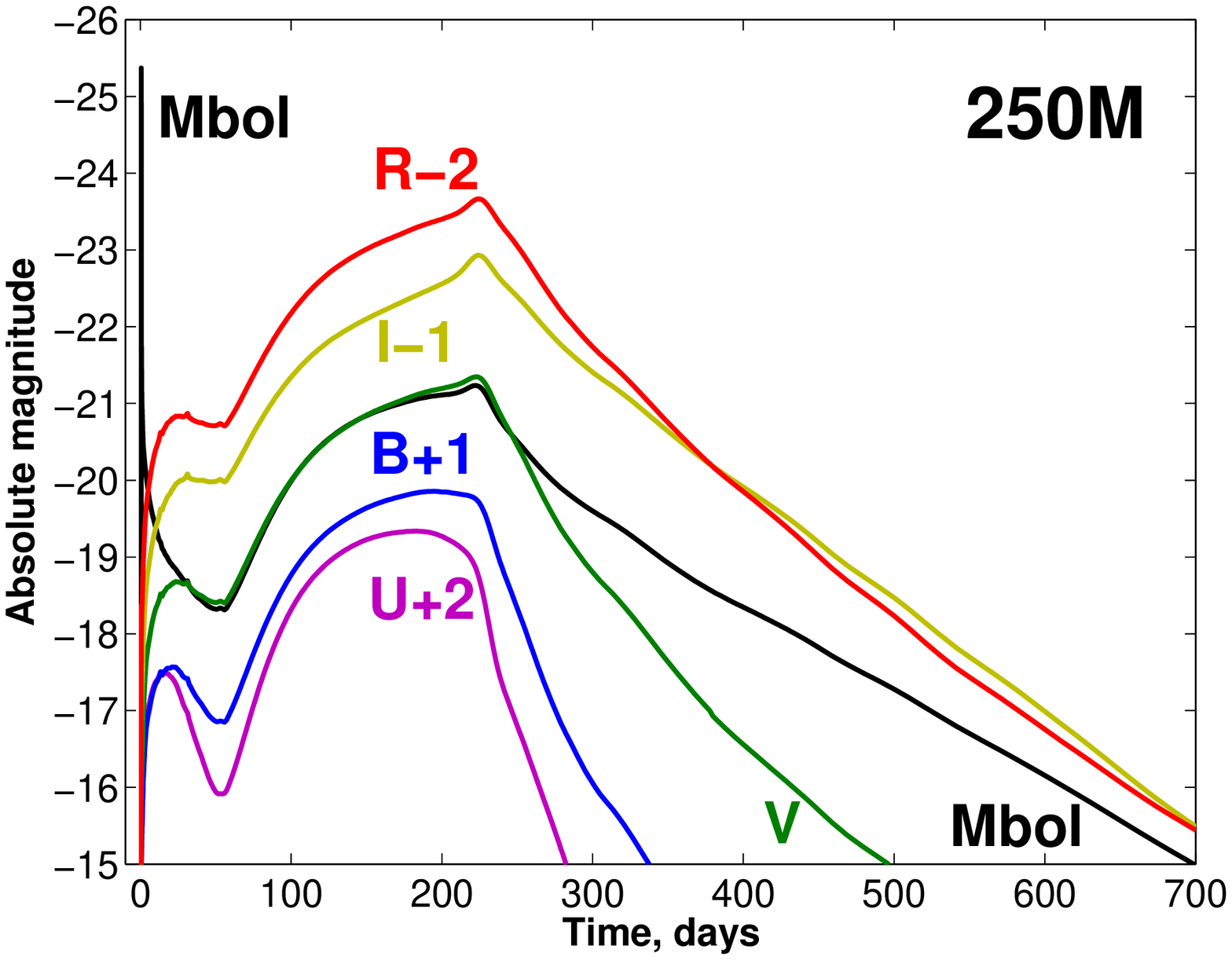}}
\caption[Bolometric and multiband (\emph{U}, \emph{B}, \emph{V}, \emph{I}, \emph{R}) light curves for 150~$M_\odot${} and 250~$M_\odot${}
PISNe.]
{Bolometric and multiband (\emph{U}, \emph{B}, \emph{V}, \emph{I}, \emph{R}) light curves for 150~$M_\odot${} and 250~$M_\odot${}
PISNe at metallicity $Z=10^{\,-3}$.  \emph{U}, \emph{B}, \emph{V}, \emph{I},
\emph{R} magnitudes are plotted with 
a shift of +2, +1, 0, -1, -2~magnitudes, respectively.}
\label{figure:bolbands}
\end{figure*}

The opacity is calculated for each frequency group taking about 160000 spectral
lines into account according to Kurucz data \citep{1991sabc.conf..441K}.
The opacity also includes photoionization, free-free absorption and
electron scattering assuming local thermodynamical equilibrium.  
Because of the large velocity gradient the opacity is
calculated accounting for the effect of an expanding medium following
\citet{1983ApJ...272..259F} and \citet{1993ApJ...412..731E}.  Local
thermodynamic equilibrium is assumed in the plasma allowing to use
the Boltzmann-Saha's distribution for ionization and level populations.  This is
needed for determining absorption, scattering and emission coefficients.  
Gamma-ray transfer is calculated using the one-group approximation for the
non-local deposition of energy from the radioactive decay
\citep{1988ApJ...325..820A}.  The code treats strong discontinuities (shock
propagation) with an artificial viscosity term.  The {\sc STELLA} code does not
follow any nuclear reactions.

To map the {\sc BEC}~models into the {\sc STELLA}~code we remeshed the models.  
The original number of zones was reduced from 1931~{\sc BEC}~zones to 242 {\sc
STELLA}~zones for the 150~$M_\odot${} model and from 2202~{\sc BEC}~zones to 
276~{\sc STELLA}~zones for the 250~$M_\odot${} model.  We reduce the number of
isotopes from 200 (from hydrogen to germanium) to 16 (H, He, C, N, O,
Ne, Na, Mg, Al, Si, S, Ar, Ca, Fe, Co, Ni).  

The envelopes of the progenitor models are optically thick and photons are not
able to leave them on a short time scale.  Therefore, {\sc STELLA}
radiation transport calculations are computationally very expensive for an input model in
which a shock is located far from the photosphere.   On the other hand, using a
more evolved input model in which the shock front is closer to the photosphere
causes more numerical instabilities just behind the shock because the {\sc
BEC}~code does not include artificial viscosity.  We therefore compared two
different cases to test how the choice of the starting model affects the final
outcome:  one in which the shock front is located at the bottom of the helium shell
and the other in which the shock is propagating in the middle of the helium
envelope.  We find that the final solutions differ only by less than 10\% in the
terms of bolometric luminosity.  In the following we use
the model in which the shock reached the bottom of hydrogen-helium envelope.  
This defines the point $t = 0$ for all the figures.

Note that the zero point in time ($t = 0$) is not the same as the time of the explosion.  
Nevertheless, the time between the onset of the pair instability explosion
(namely the highest central density) and
the beginning of the light curve calculations is 1~hour for 150~\Msun{}~model and 44~seconds for 
250~\Msun{}~model.  These are the
time intervals that the shock takes to propagate through the shallow pure helium layer to reach 
the bottom of hydrogen-helium envelope.  These intervals are relatively small 
compared to the light curve evolution (months), therefore, one can consider the 
time in all figures as approximate time since the onset of the pair instability explosion.

\section{Results}
\label{sect:results2}

\begin{figure*}
\centering
\includegraphics[width=0.47\textwidth]{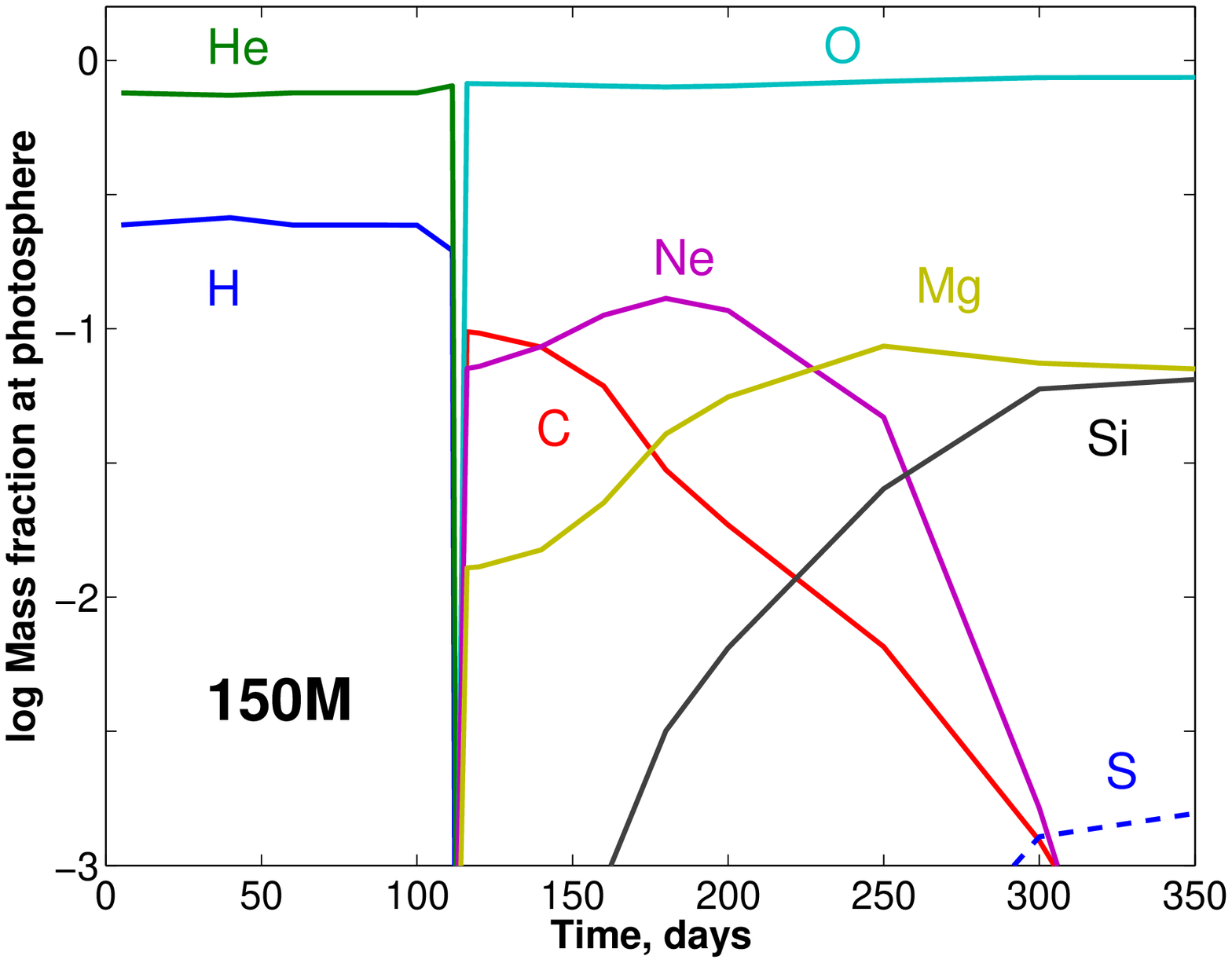}~~
\includegraphics[width=0.5\textwidth]{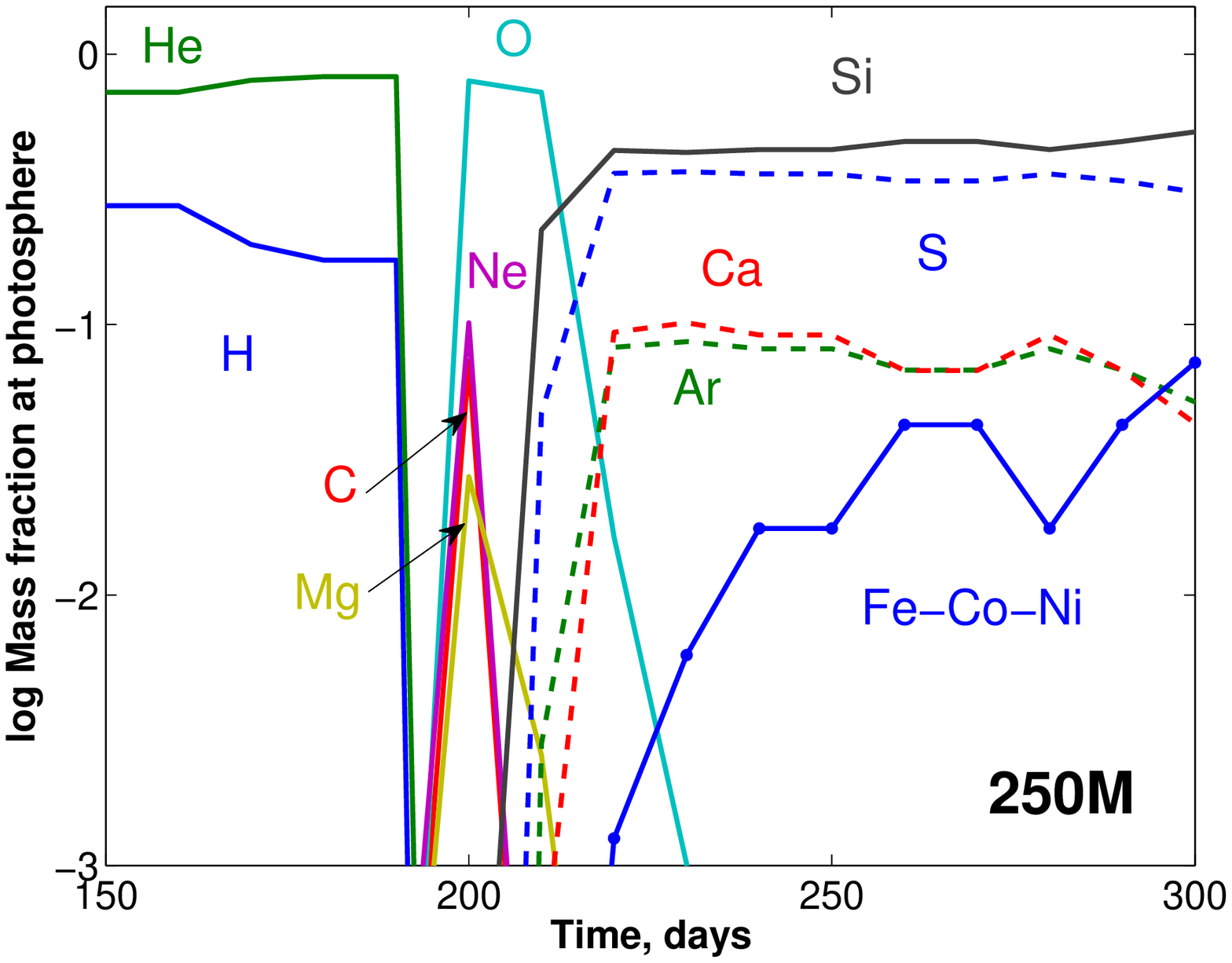}
\caption[Evolution of the composition at the receding electron-scattering photosphere for the
150~$M_\odot${} and 250~$M_\odot${} PISN models.]
{Evolution of the composition at the receding electron-scattering photosphere for the
150~$M_\odot${} and 250~$M_\odot${} PISN models.}
\label{figure:photoX}
\end{figure*}

In Figure~\ref{figure:bolbands}, we plot the bolometric and 
\emph{U}, \emph{B}, \emph{V}, \emph{R}, \emph{I} band light curves for the 150~$M_\odot${} and 250~$M_\odot${} models.  
Hereafter, in figures and tables we designate the models as Model~150M and Model~250M
respectively.  The \emph{U}, \emph{B}, \emph{V}, \emph{R}, \emph{I} magnitudes are plotted with a shift of +2,
+1, 0, -1, -2 magnitudes, respectively.  The bolometric curves follow closely
the \emph{V}-band light curves, except for early time where it
needs ultraviolet bolometric corrections, and for late time where it needs
infrared bolometric corrections \citep{2009ApJ...701..200B}.  We summarize
the details about shock breakout events (duration, maximum luminosity,
effective and color temperatures, spectral wavelength peak, spectral energy peak) 
and light curves in comparison to other radiative simulations of PISN explosion
\citep{2011ApJ...734..102K,2013MNRAS.428.3227D} in Table~\ref{table:LCchara}.

For the discussion on the light curves below, we also show the evolution
of the chemical composition at the photosphere which recedes along the
decreasing Lagrangian mass coordinate, in Figure~\ref{figure:photoX}.  The
photosphere is defined as the mass zone where the electron-scattering optical depth
in {\emph B}-band turns above 2/3.  This consideration is valid while matter
and radiation temperatures are comparable at an accuracy better than 10\%.  
The estimated period during which such conditions are present is approximately the first
350~days for Model~150M and 250~days for Model~250M.  At later times, the opacity is
dominated by line opacity and non-local thermodynamical equilibrium
consideration is required.

\begin{figure}
\centering   
\includegraphics[width=0.5\textwidth]{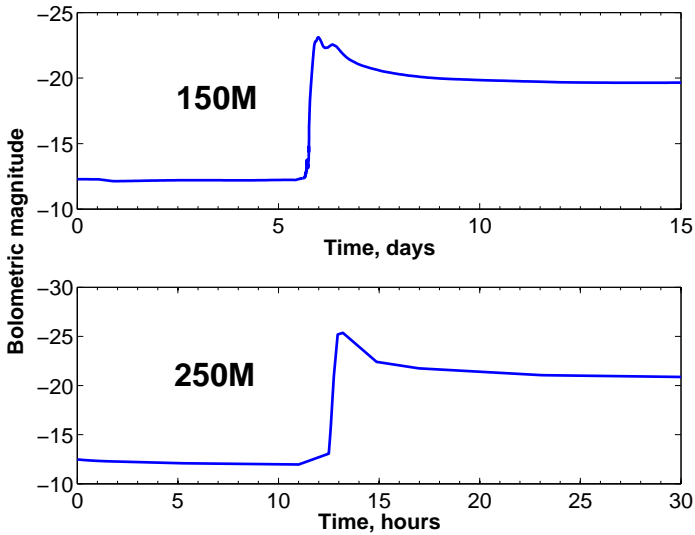}
\caption[Shock breakout events for 150~\Msun{} and 250~\Msun{} PISN explosions.]
{Shock breakout events for 150~$M_\odot${} and 250~$M_\odot${} PISN explosions.}
\label{figure:shock}
\end{figure}

Soon after the explosion, the shock wave forms when the fast expanding inner
part of the star encounters the slowly moving helium layer and the
hydrogen-helium envelope.  Because the whole envelope is optically thick it
takes a long time for photons
generated during the explosive burning to reach the surface (optically
transparent medium).  Therefore, firstly, during about 6~days for Model~150M and
12~hours for Model~250M the shock passes through the entire envelope
\citep[see e.g.][]{1987Natur.328..320S,2000ApJ...532.1132B}.  
The shock wave reaches the surface, and the envelope matter behind the shock is heated and ionized.  
When the shock emerges at the surface the energetic shock breakout flash of
ultraviolet and X-ray emission appears with a duration of a few hours.  The
shock breakout is shown in the Figure~\ref{figure:shock}.  
We draw some important conclusions about the shock breakout in Section~\ref{sect:discussion2}.

\subsection{The 150~$M_\odot${} model}
\label{subsect:150M}

Soon after the shock breakout, the temperature drops rapidly due to adiabatic cooling.  
Later the recombination losses become comparable to adiabatic cooling
\citep{1971Ap&SS..10....3G}.  The recession of the photosphere along the Lagrangian
mass coordinate is compensated by an overall expansion of the envelope.  The
combination of expansion and cooling provides the condition for only slight variations of the
luminosity for some time \citep{1976Ap&SS..44..409G,1989ASPRv...8....1I}.  This emerges as a plateau
phase in the light curve.  The plateau phase lasts for about 100~days for Model~150M.  
The light curve shape of Model~150M looks similar to that of ordinary type~IIP supernovae.  
This fact is also noticed by
\citet{2005ApJ...633.1031S} and \citet{2011ApJ...734..102K}.  
The absolute \emph{V}-band magnitude becomes about $M_V = -19$~mag at the
visual maximum, which is about 10~times brighter than an average SN~IIP.  
However, the peak luminosities of type~IIP SNe vary by a factor of 100, 
and that of our model is still contained in this range.  
Its high luminosity results from the relatively high supernova energy (about
10~foe) and the large radius (3394~\Rsun) of the progenitor.  Once the
photosphere recedes to the region below the hydrogen-rich envelope, the
luminosity decreases rapidly and then becomes governed by the radioactive decay
of $^{56}$Co. 

\begin{figure}
\centering   
\includegraphics[width=0.5\textwidth]{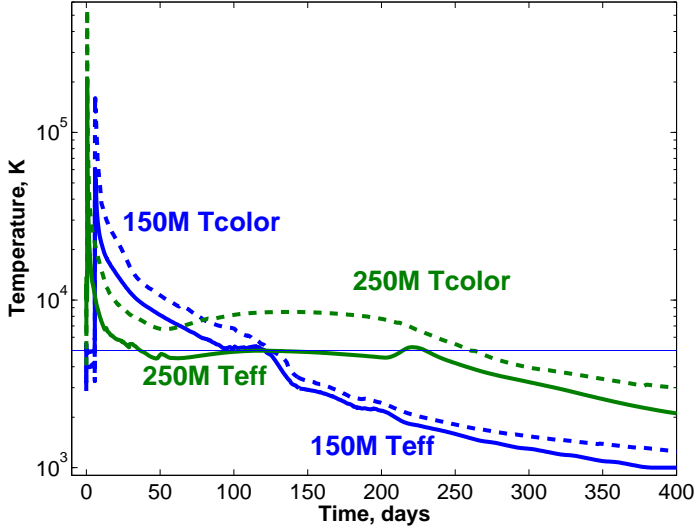}
\caption[Evolution of the effective and color temperature of 150~$M_\odot${} and 250~$M_\odot${} PISN models.]
{Evolution of the effective and color temperature of
150~$M_\odot${} and 250~$M_\odot${} PISN models.  The solid and dashed curves present effective and color temperatures 
correspondingly.  The thin line is located at 5000~K and approximately corresponds to hydrogen recombination temperature.}
\label{figure:Tcoleff}
\end{figure}

It is interesting to consider the effective temperature evolution which we present
in Figure~\ref{figure:Tcoleff}.  Note that the recombination effectively slows
the adiabatic cooling during the expansion.  The
recombination starts playing a role when recombination radiation becomes
comparable to adiabatic cooling.  This happens when the effective temperature is
approximately 10000~K \citep{1965SvA.....8..664I}, i.e. well before the
establishment of a recombination cooling wave.  The effective temperature of an
ordinary SN~II remains nearly constant ($\sim 5500$~K) after the establishment
of a recombination cooling wave.  This phase begins about 20~days after the
explosion for an ordinary SN~II
\citep{2009ApJ...701..200B,2011MNRAS.410.1739D}.  This delay of 20~days depends
mostly on the progenitor radius \citep{1971Ap&SS..10....3G,1977SvAL....3...34I}. 

Due to the very large progenitor radius of 150~$M_\odot${} PISN
(3400~$R_\odot$) the onset of the recombination cooling wave is delayed up to
day~100 in this case. For about 20~days thereafter, when the photosphere is still inside
the hydrogen-helium layer (Figure~\ref{figure:photoX}), the recombination
cooling wave is established for a while and the effective temperature is kept
at the hydrogen recombination level.  Although the mass fraction ratio between
hydrogen and helium is 4:1, the number of hydrogen atoms is about equal to the number
of helium atoms and the relative contribution to the electron density from hydrogen
stays high.  The plateau phase of this model mostly corresponds to
the phase of rapidly evolving photospheric temperature.  This is different from
the case of ordinary SN~IIP \citep[see e.g.][]{2009ApJ...701..200B}.  In
Section~\ref{subsect:compareCC} we provide the comparison of the color temperature
with those of typical and bright SNe~IIP, and with SN~2009kf which was bright in 
the near-ultraviolet (NUV) range at early time.

\subsection{The 250~$M_\odot${} model}
\label{subsect:250M}

The light curve  looks very different for our Model~250M (Figure~\ref{figure:bolbands}).  
After the shock breakout, the bolometric magnitude drops to about $M \simeq
-18$~mag during the first 50 days.  The luminosity decreases due to the
adiabatic expansion more rapidly than in Model~150M, because the progenitor of
Model~250M is more compact (i.e., $R = 745$~\Rsun{}, compared to
3394~\Rsun{} in Model~150M).  A precursor-like event happens at around day~20 in \emph{U},
\emph{B}, \emph{V}-bands at
magnitude $M \simeq -18.5$~mag.  Such precursor could be perceived as a separate 
hydrogen-rich SN being discovered long before reaching the maximum luminosity. 

A rebrightening occurs thereafter, as the energy from the
radioactive decay of nickel and cobalt in the ejecta diffuses out
\citep{2010MNRAS.405.2113D}.  The photosphere recedes below the hydrogen-rich
envelope starting at about day~175 (Figure~\ref{figure:photoX}), but unlike in the
case of Model~150M, the luminosity keeps increasing until about 220 days, reaching
$M_V = -21$~mag at the visual maximum. At later times, the light curve is
largely governed by the radioactive decay of $^{56}$Co.  A hump-like feature on
top of the main maximum phase is shown around day~220.  At this time the photosphere
leaves the oxygen-rich layer and moves down to the silicon-rich layers.  
The receding front gradually encounters a bubble of diffusing photons generated by
the radioactive decay. 


It is interesting to note that the photosphere is located in the inner layers
where there is no hydrogen and helium well before the luminosity reaches its
peak, as mentioned above (Figures~\ref{figure:bolbands} and
\ref{figure:photoX}).  Based on our results we can not
predict whether hydrogen and helium lines are expected in
the spectrum at the time of maximum luminosity due to excitation by radiation above the
photosphere.  However, \citet{2011MNRAS.410.1739D} showed that the Balmer lines
disappear after the photosphere leaves the hydrogen-rich shell of ejecta in the case of
type~II~plateau~SN and there are no Balmer-continuum photons at later time.  
The spectral models for PISNe by \citet{2013MNRAS.428.3227D} also indicate that
hydrogen lines do not appear once the photosphere moves down to the
hydrogen-free core.  Non-thermal excitation of hydrogen can happen if there is
some degree of mixing of cobalt and nickel into layers containing hydrogen
\citep{2012MNRAS.426.1671L}.  However it is expected that PISNe do not
experience such mixing during the explosion according to the recent numerical
simulations of \citet{2011ApJ...728..129J}.  Therefore, we expect that PISNe like
Model~250M will not have any hydrogen lines once the photosphere recedes to the
hydrogen-free core.  For the same reason, we do not expect helium lines in the
optical bands either, because strong mixing of helium and radioactive nickel is
needed to excite helium lines as shown in \citet{2012MNRAS.424.2139D}.  This
means that the Model~250M would appear as a type~II SN initially but look like
a SN~Ic from about day~175, well before the luminosity reaches its peak value.  

\section{Discussion}
\label{sect:discussion2}

\subsection{Comparison with other theoretical PISN light curves}
\label{subsect:comptheory}

In this section we compare our results with other PISN models from red or
blue supergiant progenitors.  These include  PISN models carried out with the
Monte Carlo radiation transport code {\sc SEDONA} (labeled `.K' in figures,
tables; \citealt{2006ApJ...651..366K,2011ApJ...734..102K}) and with the non-LTE
radiative-transfer code {\sc CMFGEN} (labeled `.D';
\citealt{2013MNRAS.428.3227D}).  Light curves from the {\sc SEDONA} simulations
are integrated from spectra provided by D.~Kasen and artificially smoothed.  
The noisy appearance is a consequence of the statistical approach of Monte Carlo
transport calculations.  The advantage of the {\sc STELLA} calculations over
the {\sc SEDONA} and the non-LTE {\sc CMFGEN} codes is the capability to simulate
the radiation-hydrodynamical evolution from the onset of the explosion.  Hence we
reproduce the emergence of the shock breakout event and the hydrodynamical
evolution during the pair instability explosion.  Both, {\sc SEDONA} and {\sc CMFGEN},
carry out the radiative transfer calculations once the homologous expansion is reached.
The codes have their advantages for spectral computations well
described in a number of papers
\citep{2006ApJ...651..366K,2007ApJ...662..487W,2010MNRAS.405.2141D,2011MNRAS.410.1739D}.

\subsubsection{Red supergiants}
\label{subsubsect:RSGs}

Red supergiant (RSG) progenitors are common for producing SNe~IIP
\citep{1971Ap&SS..10....3G,2009ARA&A..47...63S}.  Typically the light curve
from the explosion of a red supergiant is characterized by a pronounced plateau phase
lasting for about 100~days, followed by a radioactive tail \citep{1979A&A....72..287B}.  
In this section we compare our lower mass PISN Model~150M with other PISNe produced by RSG progenitors.

\begin{figure}
\centering
\includegraphics[width=0.5\textwidth]{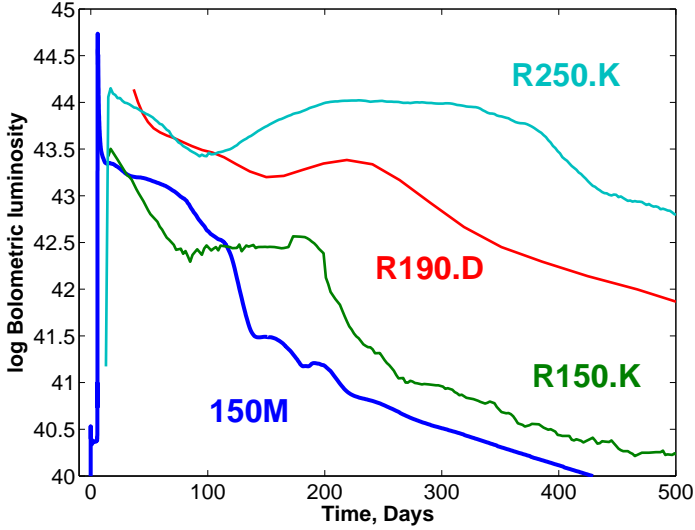}
\caption[Bolometric luminosity for the red supergiant models 150M, R150.K, R190.D, and R250.K.]
{Bolometric luminosity for the red supergiant models 150M (blue curve), R150.K (green curve), 
R190.D (red curve), and R250.K (cyan curve).}
\label{figure:RSG_lc}
\end{figure}

\begin{figure}
\centering
\includegraphics[width=0.5\textwidth]{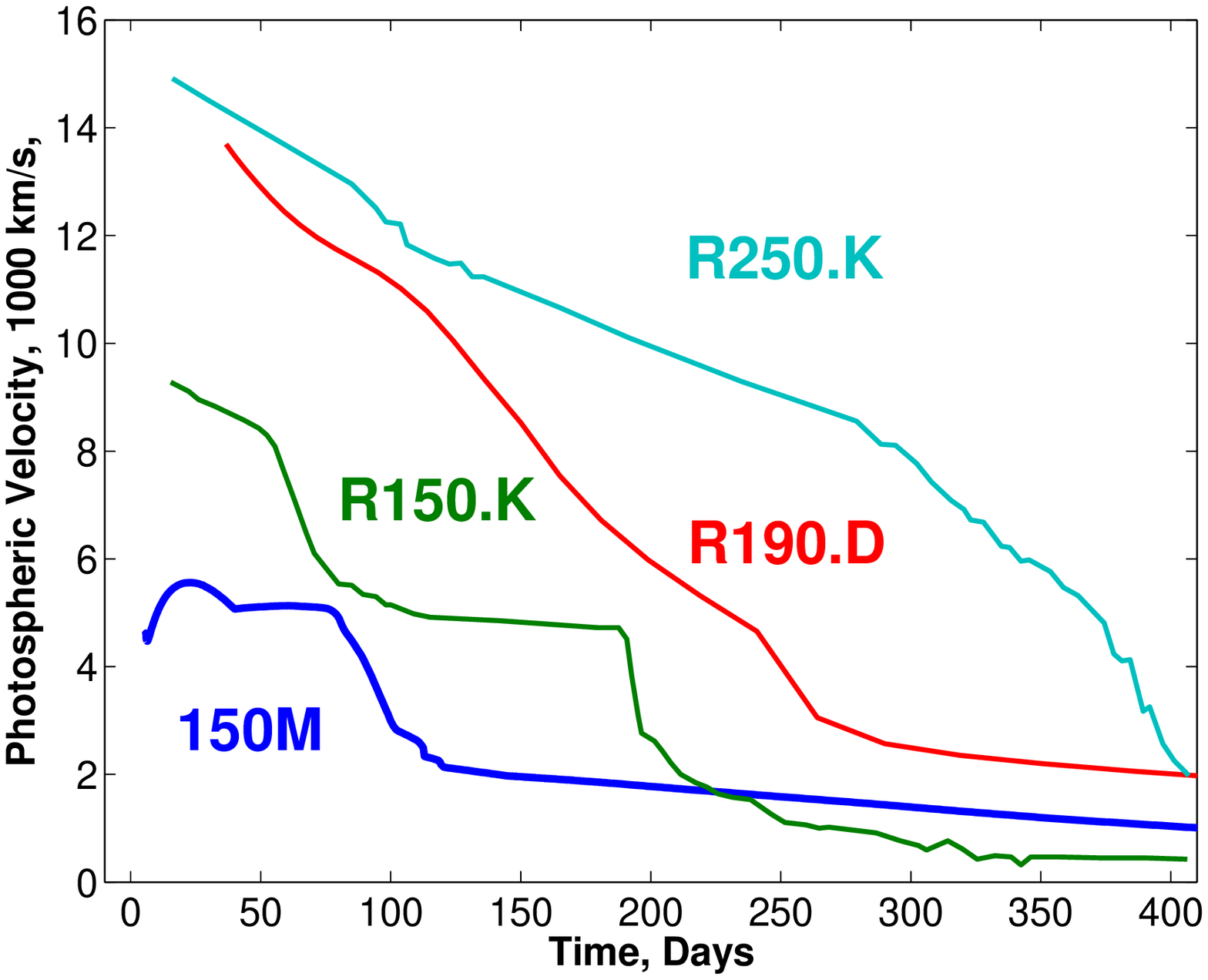}
\caption[The photospheric velocities for the red supergiant models 150M, R150.K, R190.D, and R250.K.]
{The photospheric velocities for the same models as shown in Figure~\ref{figure:RSG_lc}.  
Labels and colours have identical meanings.}
\label{figure:RSG_uph}
\end{figure}

\begin{figure}
\centering
\includegraphics[width=0.5\textwidth]{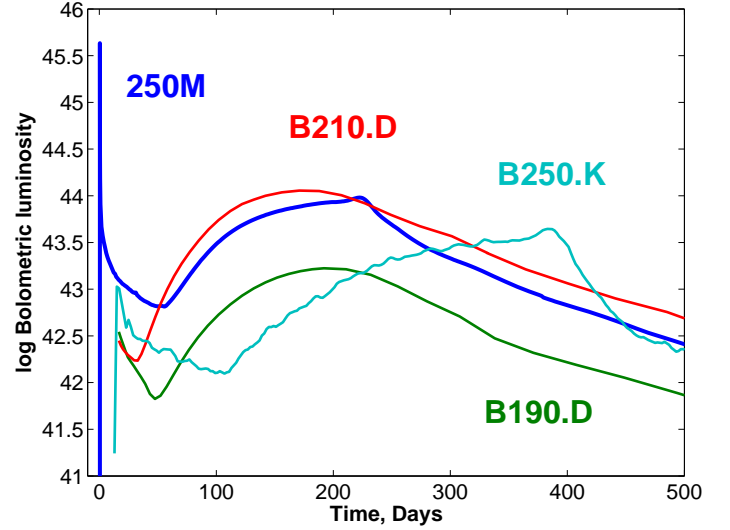}
\caption[Bolometric luminosity for the Model~250M, Model~B190.D, Model~B210.D, and Model B250.K.]
{Bolometric luminosity for the Model~250M (blue curve), Model~B190.D (green curve), Model~B210.D (red curve), 
and Model~B250.K (cyan curve).}
\label{figure:BSG_lc}
\end{figure}

In Figure~\ref{figure:RSG_lc}, we plot the theoretical light curves for different RSG
PISN models: Model~150M, Model~R150.K, Model~R150.D, and Model~R250.K (see Table~\ref{table:moddata} for details).
As already discussed by  \citet{2011ApJ...734..102K} and
\citet{2013MNRAS.428.3227D},  more massive progenitors result in higher
luminosities and broader light curves.  In particular, heating due to
radioactive decay of nickel and cobalt leads to a rebrightening in Model~R190.D and
Model~R250.K which produced large amounts of $^{56}$Ni (2.63~\Msun{} and 37.9~\Msun{},
respectively), while it is not seen in Model~R150.K and Model~150M which have only
0.07~\Msun{} and 0.04~\Msun{} of $^{56}$Ni, respectively.  The tail of each
light curve is powered by cobalt decay and its luminosity is
directly proportional to the amount of radioactive nickel generated during the
explosion.  This dependence is clearly visible in Figure~\ref{figure:RSG_lc}.

Our Model~150M is directly comparable to Model~R150.K, because it has the same
initial progenitor mass, a similar explosion energy and a similar nickel mass
(Table~\ref{table:moddata}).  Compared to Model~150M, Model~R150.K has a 500~times lower
metallicity, which results in several important differences in the progenitor
properties:
\begin{enumerate}
\item The ejecta mass is much lower for Model~150M (94~\Msun{}) than that of
Model~R150.K (143~\Msun{}).
\item The hydrogen-helium envelope masses are 29~\Msun{} in Model~150M and
71~\Msun{} in Model~R150.K, respectively.
\item Because of stronger mass loss, the
hydrogen mass fraction in the envelope of Model~150M is much lower ($X_\mathrm{H}
\approx 0.2$) than that of Model~R150.K ($X_\mathrm{H} \approx 0.7$).
\item The radius
at the pre-supernova stage is significantly larger for Model~150M ($R = 3394$~\Rsun) than
for Model~R150.K ($R = 2314$~\Rsun).
\end{enumerate}
As a consequence the overall plateau
duration is shorter in Model~150M than in Model~R150.K, because of the smaller hydrogen
envelope mass and lower mass fraction of hydrogen in the envelope.  
Moreover, nickel was additionally mixed into outer layers in Model~R150.K
which in turn caused extra nickel heating during plateau phase.

Figure~\ref{figure:RSG_uph} plots the photospheric velocities of the RSG
models. The Model~150M and Model~R150.K have a phase where the photospheric
velocity remains nearly constant, which results from the interplay between the
recession of the photospheric front and the expansion of the ejecta.  For the
models with higher $E/M$ ratio, i.e. Model~R190.D and Model~R250.K, the photospheric
velocities stay at a high level for a longer time.  This is the result of a stronger
explosion and hence a more powerful shock.  More energetic photons keep the
medium ionised for longer and the photosphere remains in hydrogen-rich regions
(larger Lagrangian mass coordinate) for a longer time (up to around day~250 and
day~300 for Model~R190.D and Model~R250.K, respectively).

The photospheric velocity in Model~150M is very low at the moment of shock breakout
(4000~km/s) because the shock spends a large fraction of its energy on ionizing
the medium while traveling along the extremely extended envelope. As showed in
\citet{1971Ap&SS..10....3G}, \citet{1977ApJS...33..515F}, \citet{2011AstL...37..194B} 
and \citet{2013MNRAS.429.3181T},
extended progenitors result in lower photospheric velocities than more compact
ones.

\begin{table*}
\caption[Shock breakout and plateau-maximum phase characteristics.]
{Shock breakout and plateau-maximum phase characteristics.  Shock 
breakout duration (defined as a full width at half-maximum), 
bolometric peak luminosity, effective and color temperature, spectral 
wavelength peak in $\AA$, spectral energy peak in keV.  
The photospheric phase begins after relaxation from the shock breakout 
and is limited by the transition to the radioactive decay tail.  
Model~150M and Model~250M are those of 150~$M_\odot${} and 250~$M_\odot${} PISNe simulated in the frame of present study.  
Labeled `R190.D', `B190.D' and `B210.D' are models from \citep{2013MNRAS.428.3227D} and 
`R150.K' and `R250.K' are models simulated by \citep{2011ApJ...734..102K}.}
\label{table:LCchara}
\begin{center}
\begin{tabular}{lccccc|cc}
\hline
\, &  \multicolumn{5}{c|}{{\bf{shock breakout}}} & \multicolumn{2}{c}{{\bf{photospheric phase}}}\\
\, &  duration, & $L$, & $T_\mathrm{eff}/T_\mathrm{color}$, &
$\lambda_\mathrm{max}$, & $E_p$, & duration, & $L$, \\
 & (hours) & (erg/s) & ($10^{\,3}$~K) & ($\AA$) & (keV) & (days) & (erg/s) \\
\hline
\textbf{150M} & 6   & $5.4\times 10^{\,44}$ &  60/160 & 170 & 0.07 & 110 & $1.6\times 10^{\,43}$            \\
R150.K        & 2   & $1.2\times 10^{\,45}$ &  90/170 & 169 & 0.07 & 200 & $3\times 10^{\,42}-10^{\,43}$      \\
R190.D        & --- & ---                   & ---     & --- & ---  & 260 & $3\times 10^{\,43}$              \\
R250.K        & 1.6 & $9.6\times 10^{\,45}$ & 130/350 & 83  & 0.15 & 410 & $6\times 10^{\,43}-10^{\,44}$      \\
\hline
B190.D        & --- & ---                   & ---     & --- & ---  & 300 & $1.6\times 10^{\,43}$            \\
B210.D        & --- & ---                   & ---     & --- & ---  & 280 & $2\times 10^{\,42}-10^{\,44}$      \\
\textbf{250M} & 1.4 & $6.2\times 10^{\,45}$ & 230/570 & 51  & 0.24 & 280 & $10^{\,43}-6\times 10^{\,43}$      \\
B250.K        & 0.3 & $1.4\times 10^{\,45}$ & 330/630 & 46  & 0.27 & 440 & $2\times 10^{\,42}-5\times 10^{\,43}$ \\
\hline
\end{tabular}
\end{center} 
\end{table*}

\subsubsection{Yellow and blue supergiants}
\label{subsubsect:BSGs}

In Figure~\ref{figure:BSG_lc}, we plot light curves for PISNe from more compact
progenitors than RSGs: Model~250M, Model~B190.D\footnote{Blue supergiant B190.D is the same
evolutionary model as red supergiant R190.D but with truncated hydrogen
atmosphere (about 30~\Msun{} of outer hydrogen layer is cut;
\citealt{2013MNRAS.428.3227D}).}, Model~B210.D, and Model~B250.K.  These PISNe produce
light curves reminiscent to that of SN~1987A: the luminosity decreases rapidly in
the beginning, but the supernova rebrightenes as the thermalized photons from the
radioactive decay of nickel and cobalt diffuse out.  The bolometric luminosity 
of our yellow supergiant Model~250M is higher than that of the other blue supergiant models during
the initial phase ($t \lesssim 50$~days), because of the larger radius of the progenitor 
\citep[e.g.][]{1993ApJ...414..712P}. 

\begin{figure}
\centering
\includegraphics[width=0.5\textwidth]{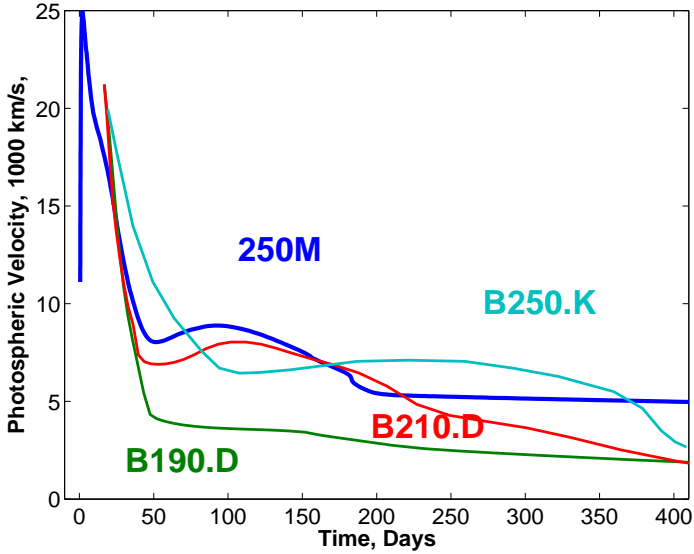}
\caption[The photospheric velocities for the Model~250M, Model~B190.D, Model~B210.D, and Model~B250.K.]
{The photospheric velocities for the same models as shown in Figure~\ref{figure:BSG_lc}.  
Labels and colours have identical meanings.}
\label{figure:BSG_uph}
\end{figure}

In Figure~\ref{figure:BSG_uph}, we show the evolution of the photospheric
velocity for the compact progenitor models.  In all of these models, the
photosphere recedes very rapidly through the outer hydrogen-helium layer after
shock breakout.  In Model~250M, Model~B210.D and Model~B250.K, where the amounts of 
nickel exceed 19~\Msun{}, the reversion of the photospheric velocity
occurs when the recombination and cooling wave encounters the expanding
``bubble'' of diffusing photons generated by nickel radioactive decay.  These
photons ionize the just recombined medium and push the photospheric front to
outer shells.  A larger envelope mass leads to a broader phase of this reverse
photosphere motion.

The sudden drop in the velocity of Model~250M occurs around day~175.  At this time
the photosphere leaves the hydrogen-helium envelope (Figure~\ref{figure:photoX})
and moves rapidly through hotter layers of oxygen, neon, carbon and magnesium
heated by diffusing nickel photons.  There is a lack of such a sharp drop in other
calculations because their input chemical structure was smeared
\citep{2011ApJ...734..102K,2013MNRAS.428.3227D} to mimic hydrodynamical mixing
happened during the explosion.  
However, it was shown by a number of studies that the degree of mixing 
for the inner regions containing radioactive nickel is not so prominent
\citep{2011ApJ...728..129J,2012ASPC..453..115C,2013ApJ...776..129C,2014arXiv1402.5960C}.  
Mixing is more efficient in the oxygen layer where the shock emerges and above it due to the propagation of the reverse shock.  
Red supergiants exhibit a higher degree of mixing compared to more compact blue supergiants, 
similar to red and blue supergiant core collapse SNe
\citep{2009ApJ...693.1780J}.

\subsection{The chemical structure during the coasting phase}
\label{subsect:coast}

\begin{figure}
\centering
\includegraphics[width=0.5\textwidth]{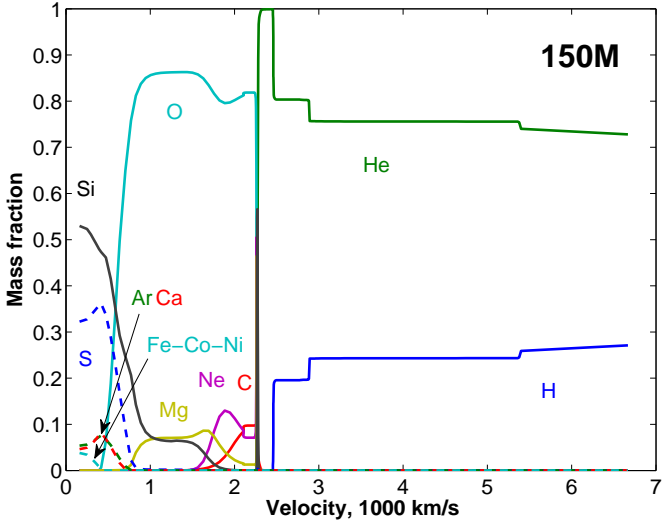}
\includegraphics[width=0.5\textwidth]{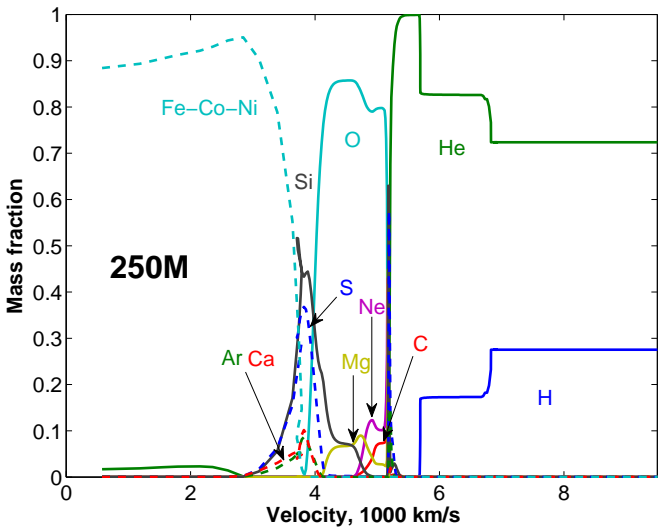}
\caption[The chemical structure of ejecta of 150~$M_\odot${} and 250~$M_\odot${} PISNe at late time.]
{The chemical structure of ejecta of 150~$M_\odot${} and 250~$M_\odot${} PISNe
at day~950 and day~1200 correspondingly.}
\label{figure:coastingX}
\end{figure}
 
Figure~\ref{figure:coastingX} shows the ejecta structure of Model~150M and Model~250M
at coasting phase.  In fact, the coasting phase begins around day~10 for both our models.  
In Figure~\ref{figure:coastingX}, we plot the chemical composition of the
ejecta at day~950 and day~1200, respectively.  The plots demonstrate 
what would be the degree of Doppler broadening for the spectral lines 
of the given elements 10~days after the explosion and later.

The oxygen shell in Model~150M, which contains small amounts of carbon, neon,
magnesium and silicon, expands at a velocity of 1000~--~2200~km~s$^{-1}$.  The inner silicon-sulfur shell  
moves at velocities below 2000~km~s$^{-1}$.  Model~250M 
shows systematically higher velocities.  The oxygen-rich layers travel at
4000~--~5000~km~s$^{-1}$, the silicon-sulfur shell --- at about 4000~km~s$^{-1}$, and 
the nickel-rich layer moves at 0~--~3900~km~s$^{-1}$. 

In the Model~250M the structure has some peculiarity at the sulfur-silicon layer
moving at velocity around 4000~km~s$^{-1}$.  The reason is the hydrodynamical
effect of nickel heating
\citep{1988ApJ...331..377A,1988ApJ...330..218W,1989ApJ...341L..63A,2000ApJ...530..744P,2007ApJ...662..487W}.  
The additional radiation pressure produced by nickel heating provides
an acceleration of the innermost 30~\Msun{} of ejecta.  The effect increases
the velocity field in the inner part of the ejecta by a few hundred km~s$^{-1}$ and
decreases the density.  The overall changes in density and velocity
do not exceed 10\% on a relative scale.  This
effect has no influence on the light curve shape around the luminosity peak
\citep{2007ApJ...662..487W} because the photosphere retains far from the innermost
region at this time.


\subsection{Comparison with observed SNe}
\label{subsect:compareCC}

\subsubsection{Relatively low-mass PISNe and type~IIP SNe}
\label{subsubsect:compareCC1}

\begin{figure}
\centering
\includegraphics[width=0.5\textwidth]{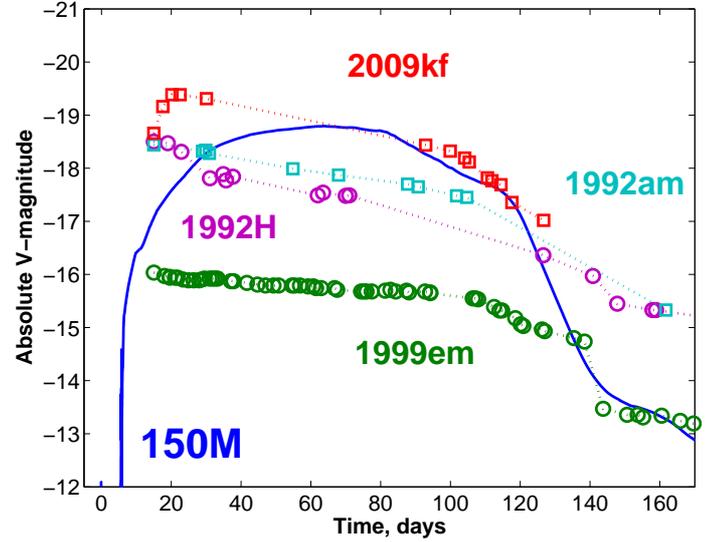}
\caption[The absolute \emph{V}-band light curve for 150~$M_\odot${} PISN is shown together with the absolute
\emph{V}-band magnitudes for typical and bright SNe~IIP.]
{The absolute \emph{V}-band light curve for 150~$M_\odot$ PISN (blue solid line) is shown together with the absolute
\emph{V}-band magnitudes for typical plateau SN~1999em (green circles) and 
three bright plateau SNe 1992H (magenta circles), 1992am (cyan squares), 2009kf
(red squares).}
\label{figure:plato}
\end{figure}

Our Model~150M is particularly relevant for the identification 
of PISNe in the local Universe.  This is because SNe from the low-mass
end of the PISN regime like Model~150M are expected to be more abundant than those from the
high-mass end like our Model~250M.  In Figure~\ref{figure:plato}, therefore,
we show the first 170~days of light curves for our synthetic PISN models and
compare it to several observed plateau SNe. 
The data for these particular SNe are taken from the Sternberg Astronomical
Institute Supernova Light Curve Catalogue
\citep{SAIMSU..SNeLCs..cat,SAIITEP..SNeLCs..cat} and are compiled from the original data
\citep{1994AstL...20..374T,1996AJ....111.1286C,2001PhDT.......173H,2010ApJ...717L..52B}.

\begin{figure}
\centering    
\includegraphics[width=0.5\textwidth]{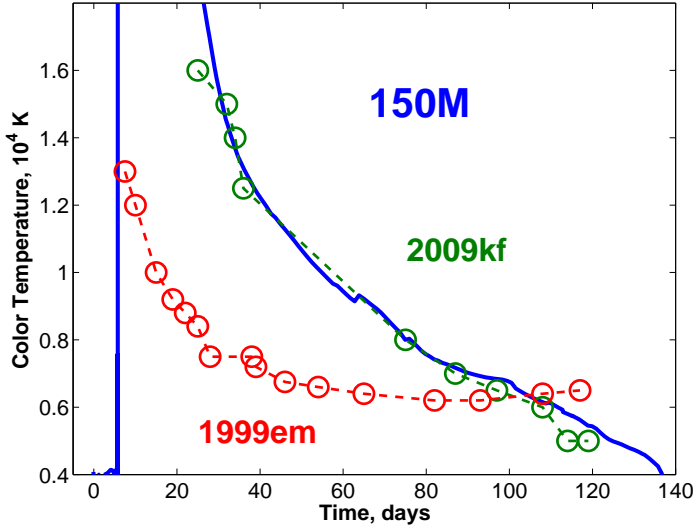}
\caption[Color temperature evolution of 150~$M_\odot${} PISN is shown together with those of
typical SN~IIP SN~1999em and NUV-bright SN~2009kf.]
{Color temperature evolution of 150~$M_\odot${} PISN is shown together with those of
typical SN~IIP SN~1999em and NUV-bright SN~2009kf. SN~1999em and SN~2009kf data are taken from 
\citet{2009ApJ...701..200B,2010ApJ...717L..52B}. SN~2009kf data are shifted in time by 15~days.}
\label{figure:Tcompar}
\end{figure}

Model~150M has a plateau phase during the first 115~days which is not unusual for SNe~IIP.  
Although this model has a higher envelope mass (29~\Msun{})
than those of typical SNe~IIP progenitors ($ < 10$~\Msun{}), the helium mass
fraction is very high (about 80\%).
At the same time it is expected that the envelopes of
typical SN~IIP progenitors have helium mass fractions of 35\%~--~50\%
depending on their initial masses \citep{2002RvMP...74.1015W,2012ARA&A..50..107L}.  Helium recombines at
higher temperatures to reduce the electron scattering opacity and the hydrogen
recombination front recedes more rapidly with a higher fraction of helium in
the envelope~\citep{2009ApJ...703.2205K}.  This makes our Model~150M to have a fairly
short duration of the plateau phase compared to high redshift RSG PISN models (cf. Figure~\ref{figure:RSG_lc}).  
The plateau duration of the light curve for our Model~150M is comparable to those of 
typical SNe~IIP, despite the relatively high envelope mass. 

During the plateau phase, the \emph{V}-band magnitude varies by 1-2~magnitudes
\citep{1979A&A....72..287B,2003ApJ...582..905H}.  Compared
to the typical plateau supernova SN~1999em, the plateau luminosity of the
Model~150M is higher by 2-3~magnitudes, but comparable to those of the three
bright SNe~IIP (SN~1992H, 1992am and 2009kf, cf. Figure~\ref{figure:plato}).  The estimated nickel masses for
these SNe~IIP are comparable or higher than that in Model~150M: $M_{\mathrm{Ni}} = 0.058~M_\odot${}, $0.075 M_\odot${}, $0.36
M_\odot${} for 1999em, 1992H, 1992am
\citep{1996AJ....111.1286C,2003ApJ...582..905H,2003MNRAS.346...97N,2011ApJ...729...61B},
respectively, and limited by 0.4~\Msun{} for 2009kf \citep{2010ApJ...717L..52B}. 

Our Model~150M has a very large initial radius of 3394~$R_\odot${} (see
Table~\ref{table:moddata}),  while those of ordinary SNe~IIP progenitors have radii of
less than 1000~$R_\odot${}.  Due to this difference our PISN explosion and SN~IIP
explosions have a different appearance of the shock breakout event.  The shock breakout
duration is longer for a larger progenitor \citep{2011ApJS..193...20T}.  The color
temperature is higher for a SN~IIP shock breakout than that for our PISN
($T_\mathrm{col}\sim R^{\,-1/2}${}) while the peak luminosity is similar depending
mostly on the explosion energy \citep{2013MNRAS.429.3181T}.  Since our PISN shock
breakout is redder the spectral maximum occurs at a longer
wavelength (see Table~\ref{table:LCchara}).  Unfortunately the shock breakout
of local SNe~II is very difficult to detect because it appears as a
ultraviolet/X-ray burst lasting only minutes to hours \citep{2004MNRAS.351..694C}.

The large progenitor radius of our 150~\Msun{}~PISN model has consequences for the
photospheric temperature evolution.  In Figure~\ref{figure:Tcompar}, we compare
the color temperature of our 150~\Msun{}~PISN with those of the typical SN~IIP
SN~1999em and the near UV-bright SN~2009kf.  The color temperature evolution of our model
is very different to that of SN~IIP 1999em, but similar to
that of SN~IIP SN~2009kf, which had a high NUV~excess at early
time.  Nevertheless, the NUV light curve of our model does not show such
a high luminosity as SN~2009kf.  The maximum NUV luminosity reaches
$-20.5$~mag and $-22$~mag for our 150~\Msun{}~PISN model and for SN~2009kf, respectively
\citep{2010ApJ...717L..52B}.  The high NUV~luminosity of 2009kf is explained by
the interaction of the SN~shock with a dense stellar wind preceding the SN~explosion
\citep{2011MNRAS.415..199M}. 
The higher temperature is the direct consequence of the
ultraviolet excess.  However, in case of our low mass PISN model the shock breakout event
occurs and the high temperature is related to the relaxation of the highly excited medium.  
Any interaction of the SN ejecta with the progenitor wind is neglected in our model.

\begin{figure}
\centering
\includegraphics[width=0.5\textwidth]{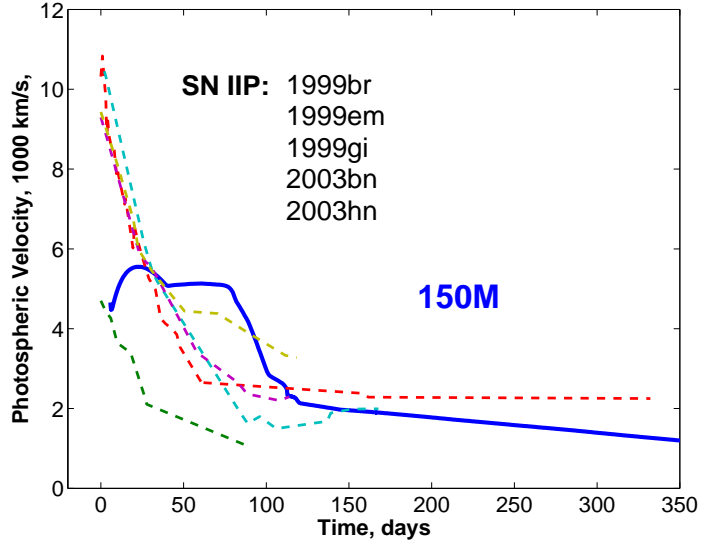}
\caption[The photospheric velocity of our 150~$M_\odot${}~PISN model and observed SNe~IIP.]
{The photospheric velocity for our 150~$M_\odot${}~PISN model (blue solid line) is shown together 
with observational data for several SNe~IIP
\citep{Jones:2008tz}.  The zero point for the observed data is the first spectroscopic observation.}
\label{figure:uph}
\end{figure}

Another possible way to distinguish a SN~IIP from a PISN explosion is to check the
photospheric velocities.  In Figure~\ref{figure:uph} we show the photospheric
velocity for our 150~$M_\odot${} PISN model along with those of a few ordinary plateau SNe taken from
\citet{Jones:2008tz}.  The estimate for the photospheric velocity is based on the
H$_\beta$ absorption line \citep{2009ApJ...696.1176J}.  The distinct property
of our model is the low photospheric velocity at earlier time compared to the
maximum photospheric velocities of SNe~IIP
\citep[cf.,][]{1971Ap&SS..10....3G,2004ApJ...617.1233Y}.  
The photospheric velocities at later time are similar for both
SNe~IIP and our low mass PISN model.

\subsubsection{SLSNe linked to type Ic SNe}
\label{subsubsect:2007bi}

\begin{figure*}
\centering
\includegraphics[width=0.5\textwidth]{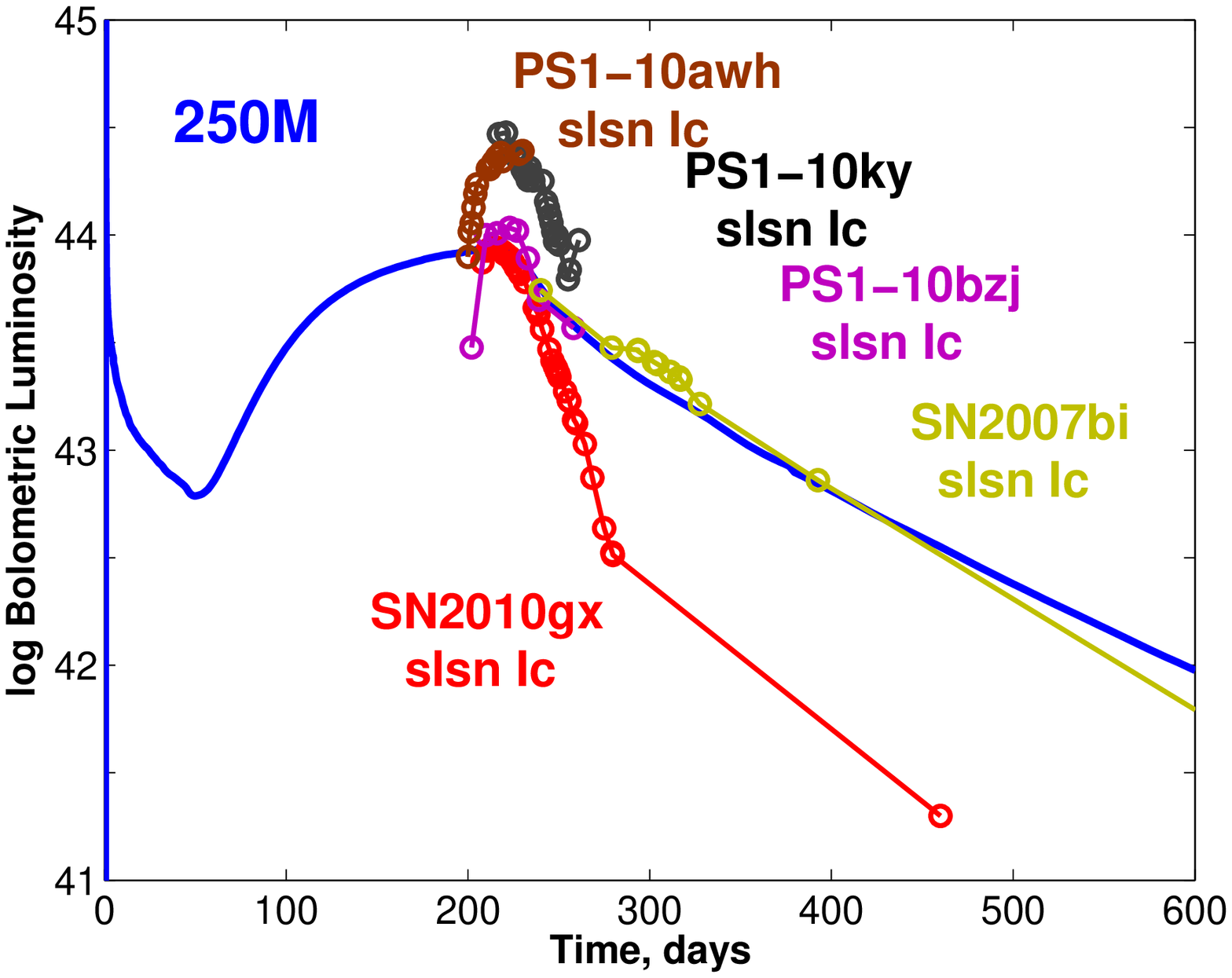}~~
\includegraphics[width=0.51\textwidth]{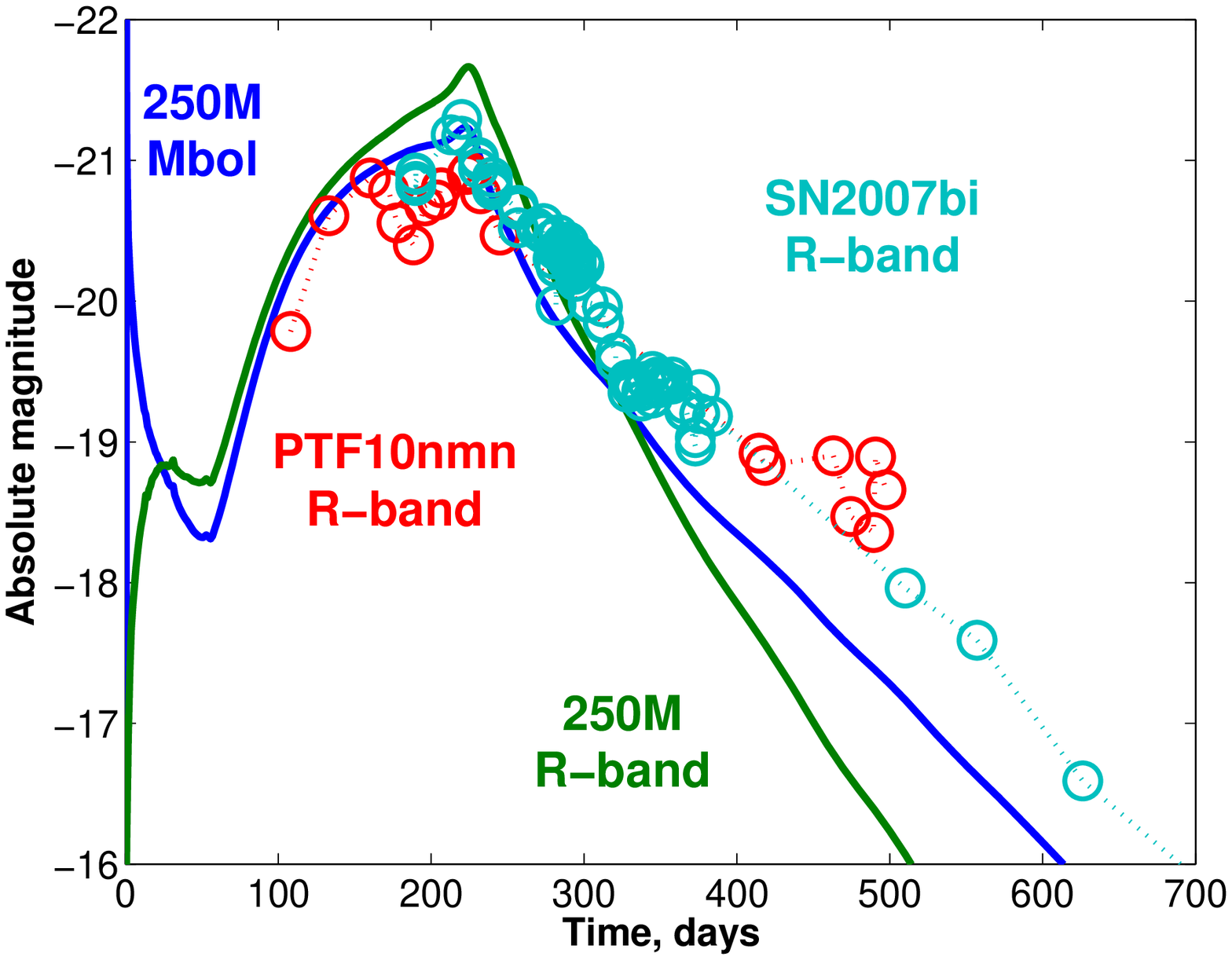}
\caption[The absolute bolometric light curve for our 250~$M_\odot$~PISN model and for SLSN~Ic.]
{
{\emph{Left panel}}:
The absolute bolometric light curve for our 250~$M_\odot$~PISN model (blue solid curve) is shown together with 
quasi-bolometric and bolometric light curves 
of superluminous type~Ic SN~2010gx (red, \citealt{2013ApJ...770..128I}),
PS1-10awh and PS1-10ky (green and black, \citealt{2011ApJ...743..114C})
PS1-10bzj (magenta, \citealt{2013ApJ...771...97L}),
and PISN candidate SN~2007bi (yellow, \citealt{2010A&A...512A..70Y}).\\
{\emph{Right panel}}:
The absolute R-band and bolometric light curve for our 250~$M_\odot$~PISN model (green and blue) with
superimposed R-band data for SLSN 2007bi (cyan) and PTF-10nmn (red). Data are taken from \citet{2012IAUS..279..253G}.\\
For both plots the observed curves are shifted in time by 200~days to coincide with the maximum phase of the theoretical curve.}
\label{figure:snIbc}
\end{figure*}

As discussed in Section~\ref{sect:results2} the photosphere of our 250~$M_\odot${}
PISN model during its peak luminosity phase recedes to the bottom of the
hydrogen-rich envelope \citep[see also][]{2011ApJ...734..102K,2013MNRAS.428.3227D}.  Therefore, one could
classify it as a SN~Ic, like SN~2007bi, if it was discovered during the maximum
or post-maximum phase.  In Figure~\ref{figure:snIbc}, therefore, we compare
the synthetic bolometric light curve of our 250~$M_\odot${} model with the quasi-bolometric 
light curves of some superluminous type~Ic SNe: SN~2010gx,
PS1-10awh, PS1-10ky PS1-10bzj, SN~2007bi, and PTF10nmn.

From Figure~\ref{figure:snIbc} (left panel) it is clear that Model~250M 
can not explain several of the unusually luminous SNe~Ic, because of the very broad light
curve of our model.  
Unambiguous evidence for a pair instability explosion of an initially high mass
progenitor would be the observation of an extremely long rise to the maximum phase
\citep{2013arXiv1310.1311B}.  Therefore, some attempts are
made to complete the light curves with retrospective detection of data points
before the maximum phase\citep{2013Natur.502..346N}.  The synthetic light curve of our 250~$M_\odot${}~PISN model shows
200~days of rise while many SNe~Ibc and SLSNe demonstrate significantly shorter
rise time less than 40~days \citep[see
e.g.][]{2009ApJ...702..226M,2011ApJ...741...97D,2011ApJ...743..114C}.  A good
example of a SLSN with a reliable long-lasting rise is PTF10nmn.  The data for this
particular SLSN and for SN~2007bi are shown in the right panel of
Figure~\ref{figure:snIbc}.  Our Model~250M agrees well with the broad light curves of PTF10nmn and SN~2007bi.

\begin{figure}
\centering
\includegraphics[width=0.5\textwidth]{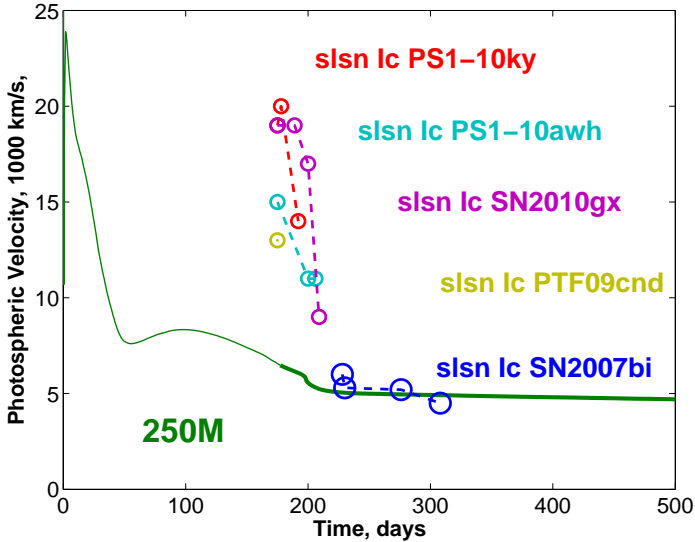}
\caption[The photospheric velocity of our 250~$M_\odot${}~PISN model and of
several SLSN~Ic.]
{The photospheric velocity of our 250~$M_\odot${}~PISN model and of
several SLSN~Ic.  The observed data are taken from
\citet{2010A&A...512A..70Y,2011ApJ...743..114C}.  SN~2007bi points show the lower velocity limit
measured from O~I~$\lambda~7774$ which we use as a best estimate for the photospheric velocity.  
The observed data are shifted to the maximum phase of the theoretical curve (day~175).  
The bold part of the theoretical curve (green) covers the maximum phase (from day~175) and
the successive decline phase.}
\label{figure:uphsnIc}
\end{figure}

The photospheric velocity of our high mass PISN model around the luminosity peak is
smaller than typical velocities of
luminous SNe~Ibc (Figure~\ref{figure:uphsnIc}).  The reason for the low velocity is
the high ejecta mass of our model.  This may be another criterion for distinguishing luminous
PISNe powered by large amounts of nickel from superluminous CCSNe.  
However, the photospheric velocities measured for SN~2007bi demonstrate that
the ejecta of this particular SLSN moves at a lower velocity than those of other
SLSNe, which is in good agreement with our model.  This
renders precise spectroscopic observations important to shed light on this
question.


Being discovered around or after its maximum, SN~2007bi resembles some 
other SLSNe which show a short rise to their peak luminosity \citep{2013Natur.502..346N}.  Particularly,
this rules out the pair instability origin of these SNe.  Nevertheless,
we conclude that our Model~250M agrees with observed properties (light curve, photospheric velocity) 
of SLSN~2007bi well.  Therefore, SN~2007bi might emerge from pair instability
explosion of very massive star with initial mass above 200~\Msun{}.

\section{Conclusions}
\label{sect:conclusions2}

We carried out simulations of shock breakouts and light curves of
pair instability supernovae using two evolutionary models of 150~$M_\odot${} and
250~$M_\odot${} at metallicity $Z=10^{\,-3}$
\citep{2007A&A...475L..19L,2014paper1..K}.  We used the radiation hydrodynamics
code {\sc STELLA} for this purpose \citep{2006A&A...453..229B}.  The considered
metallicity ($Z = 10^{\,-3}$) is among the highest of PISN models that have
been so far presented in the literature \citep{1990A&A...233..462H,2013arXiv1312.5360W}.  
Therefore, our models may serve as useful
references for future studies on PISNe observed in the local Universe, as well
as in the early Universe.

From our qualitative comparison to ordinary core collapse SNe we conclude
that it is difficult to distinguish low mass pair instability
explosions from hydrogen-rich core collapse explosions.  The photometric and spectroscopic
observations, including X-ray and ultraviolet (for detection of shock breakout events), 
should be very detailed from the earliest epoch to help shedding light on this.  
The increasing number of SN surveys allows to increase the
number of discovered SNe and detailed data from the very
early epoch after explosion, especially those missions which have the short cadences (e.g. PTF).

Given the low-mass preference of the stellar initial mass function, a large fraction of PISNe that
will be observed in the local Universe could resemble  our 150~$M_\odot${} model,
which represents PISNe from the low-mass end of the PISN regime.  
These PISNe are predicted to have the following characteristics:
\begin{enumerate}
\item The progenitors are likely to be red-supergiants having very extended envelopes ($R\sim 3000~\mathrm{R_\odot}${}), 
if they can retain some fraction of the hydrogen envelopes by the time of explosion.  Our 150~$M_\odot${} model has the final mass of
94~$M_\odot${}
and the envelope mass is 29~$M_\odot${}, which is significantly smaller than in the corresponding case of zero or extremely low metallicity ($\sim
70~M_\odot${}).  
The hydrogen mass fraction in the envelope is only about 0.25.
\item The resulting PISN would appear to be a bright type~IIP supernova like SN~2009kf.  
Its luminosity at the visual maximum would be typically higher by 2-3~magnitudes than average SNe~IIP, 
although the total amount of radioactive nickel would be more or less similar to those from usual
hydrogen-rich core-collapse supernovae ($\sim 0.05~M_\odot${}), depending 
on the final mass of the progenitor.
\item The plateau duration would be similar to those of ordinary SNe~IIP, but much shorter than in the corresponding case at extremely low metallicity 
because of the relatively low mass of the envelope and the low hydrogen mass fraction.  
\item The shock breakout duration would be somewhat longer ($\sim 6$~hrs) and redder ($0.07$~keV) than those of ordinary SNe~IIP. 
\item The photospheric color temperature would be systematically higher than those of ordinary SNe~IIP, and its evolution would look quite similar
to that of SN~2009kf, which is an unusually bright SN~IIP with a NUV-excess.
\item Because of the very large radius of the progenitor, the photospheric velocity at early times would be systematically lower than those of ordinary
SNe~IIP (Figure~\ref{figure:uph}). 
\end{enumerate}

We also conclude that some observed luminous SNe~Ic could have emerged from a pair instability explosion.  
Careful and deep photometric and spectroscopic observations would help to 
differentiate a pair instability explosion from SN~Ic, in particular for the rise epoch and the tail.  
A general property of PISN explosions from the high mass regime is a slow light curve evolution due to
massive ejecta.  This causes a long rise to the peak
luminosity and a long transition to the radioactive tail.

It was previously noted \citep{2005ApJ...633.1031S,2011ApJ...734..102K,2013ApJ...777..110W,
2013MNRAS.428.3227D} that a PISN from the high mass end of
the PISN regime does not resemble any of the observed supernovae so far. However, we
demonstrated that SLSNe~2007bi and PTF10mnm may fit well to our high mass
PISN Model~250M.  This concerns the light curve shape, the peak luminosity,
the photospheric velocity and the bulk ejecta masses.  

We suggest the following criteria to distinguish high mass PISN from CCSN:
\begin{enumerate}
\item A short precursor in \emph{U}, \emph{B}, \emph{V}-bands at about $-19$~mag lasting less than 40~days which can appear itself as
a SN long before (e.g. half a year -- a year before) the main maximum.
\item The pronounced rise time is larger than 200~days, which is significantly longer than for
ordinary SN~Ic.
\item A PISN may evolve from hydrogen-rich to hydrogen-poor type.
\item The nebular luminosity is powered by radioactive nickel decay and determined by the amount of produced nickel.  
Large amounts of radioactive nickel, tens of solar masses, produced in PISN significantly exceed typical
0.05~--~0.5~$M_\odot${} of nickel left by ordinary SNe~Ic.
\item The photospheric velocity is lower than the velocities of SNe~Ic during the whole evolution.  
On top of that the PISN photospheric velocity has a peculiar evolution during the rise to maximum
light.
\end{enumerate}

Increasing SN statistics allows to discover the
brightest SNe together with others \citep{2002SPIE.4836..154K,2008arXiv0805.2366I,2009PASP..121.1395L}.  
According to \citet{2007A&A...475L..19L} one pair instability explosion in
the local Universe occurs among one thousand
core collapse SNe.  At present the number of discovered SNe per year
surpasses one thousand, therefore, we expect several PISNe among
the large number of discovered SNe.
However, their unambiguous
identification may be challenging, which we hope to facilitate with 
our present study.

\begin{acknowledgements}

We thank Stephen Smartt, Matt Nicholl, Luc Dessart, and Takashi Moriya for fruitful and useful
discussions, Nikolay Pavlyuk for providing observed light curves, 
Mat\'{i}as Jones, Cosimo Inserra, Maria-Teresa Botticella for providing the
observed data, Daniel Kasen and Luc Dessart who provided the data from
their simulations.  
SB acknowledges the support by the grant of the Russian Federation
government 11.G34.31.0047 and also partial support by grants for
Scientific Schools 5440.2012.2, 3205.2012.2, and joint RFBR-JSPS grant
13--02--92119.

\end{acknowledgements}

\bibliographystyle{aa}
\bibliography{../references}

\end{document}